\documentstyle[12pt]{article}   

\large
\newcommand{\beq}{\begin{equation}}
\newcommand{\eeq}{\end{equation}}

\makeatletter
\@addtoreset{equation}{section}
\makeatother

\newtheorem{Theorem}{Theorem}[section]
\newtheorem{Definition}{Definition}[section]
\newtheorem{Lemma}{Lemma}[section]
\newtheorem{Corollary}{Corollary}[section]
\def\un{\underline}

\def\be{\begin{equation}}
\def\ee{\end{equation}}
\def\ba{\begin{eqnarray}}
\def\ea{\end{eqnarray}}

\def\A{{\cal A}}

\def\agb{{\overline {{\cal A}/{\cal G}}}}

\def\Ab{{\overline \A}}

\def\Comp{{\mathchoice
{\setbox0=\hbox{$\displaystyle\rm C$}\hbox{\hbox to0pt
{\kern0.4\wd0\vrule height0.9\ht0\hss}\box0}}
{\setbox0=\hbox{$\textstyle\rm C$}\hbox{\hbox to0pt
{\kern0.4\wd0\vrule height0.9\ht0\hss}\box0}}
{\setbox0=\hbox{$\scriptstyle\rm C$}\hbox{\hbox to0pt
{\kern0.4\wd0\vrule height0.9\ht0\hss}\box0}}
{\setbox0=\hbox{$\scriptscriptstyle\rm C$}\hbox{\hbox to0pt
{\kern0.4\wd0\vrule height0.9\ht0\hss}\box0}}}}
\def\Co{{\mathchoice
{\setbox0=\hbox{$\displaystyle\rm C$}\hbox{\hbox to0pt
{\kern0.4\wd0\vrule height0.9\ht0\hss}\box0}}
{\setbox0=\hbox{$\textstyle\rm C$}\hbox{\hbox to0pt
{\kern0.4\wd0\vrule height0.9\ht0\hss}\box0}}
{\setbox0=\hbox{$\scriptstyle\rm C$}\hbox{\hbox to0pt
{\kern0.4\wd0\vrule height0.9\ht0\hss}\box0}}
{\setbox0=\hbox{$\scriptscriptstyle\rm C$}\hbox{\hbox to0pt
{\kern0.4\wd0\vrule height0.9\ht0\hss}\box0}}}}
\def\Rl{{\mathchoice
{\setbox0=\hbox{$\displaystyle\rm R$}\hbox{\hbox to0pt
{\kern0.4\wd0\vrule height0.9\ht0\hss}\box0}}
{\setbox0=\hbox{$\textstyle\rm R$}\hbox{\hbox to0pt
{\kern0.4\wd0\vrule height0.9\ht0\hss}\box0}}
{\setbox0=\hbox{$\scriptstyle\rm R$}\hbox{\hbox to0pt
{\kern0.4\wd0\vrule height0.9\ht0\hss}\box0}}
{\setbox0=\hbox{$\scriptscriptstyle\rm R$}\hbox{\hbox to0pt
{\kern0.4\wd0\vrule height0.9\ht0\hss}\box0}}}}

\def\un{\underline}

\title{Kinematical Hilbert Spaces for Fermionic and Higgs Quantum Field
Theories}
\author{T. Thiemann\thanks{thiemann@math.harvard.edu}
\thanks{New Address : Albert-Einstein-Institut,
Max-Planck-Institut f\"ur Gravitationsphysik, Schlaatzweg 1, 14473 
Potsdam, Germany, Internet : thiemann@aei-potsdam.mpg.de}\\
       Physics Department, Harvard University, \\
       Cambridge, MA 02138, USA}
\date{{\small \today \\  Preprint HUTMP-97/B-364}}

\begin{document}

\maketitle

\begin{abstract}
We extend the recently developed kinematical framework 
for diffeomorphism invariant theories of connections for compact 
gauge groups to the case of a diffeomorphism invariant quantum field theory
which includes besides connections also fermions and Higgs fields. This
framework is appropriate for coupling matter to quantum gravity.

The presence of diffeomorphism invariance forces us to choose a   
representation which is a rather
non-Fock-like one : the elementary excitations of the connection are along
open or closed strings while those of the fermions or Higgs fields are at 
the end points of the string. 

Nevertheless we are able to promote the classical reality conditions 
to quantum adjointness relations which in turn uniquely fixes the 
gauge and diffeomorphism invariant probability measure that underlies the 
Hilbert space. 

Most of the fermionic part of this work is independent of the  
recent preprint by Baez and Krasnov and earlier work by Rovelli and 
Morales-Tec\'otl because we use new canonical fermionic variables,
so-called Grassman-valued half-densities, which enable us to 
to solve the difficult fermionic adjointness relations.
\end{abstract}

\section{Introduction}

A fair amount of intuition about the kinematical structure of quantum 
field theories comes from free, scalar quantum field theories in a 
Minkowski background. 
In this paper we will see that this intuition coming from free scalar
field theories is misguiding once we give up 
one or both of the following two essential ingredients of this field 
theory :\\ 
1) The quantum configuration space is a vector space,\\
2) There is available, a fixed (Minkowski) background metric.\\
More concretely, we have the following :\\
The Osterwalder-Scrader axioms of constructive quantum field theory
\cite{Jaffe}
are first of all only appropriate for kinematically linear theories,
that is, the space of quantum fields is a vector space (before imposing 
the quantum dynamics). This follows from the fact that right from the 
beginning the measure on the space of quantum fields is required to be 
supported 
on the space of tempered distributions ${\cal S}'$ on $\Rl^{d+1}$ where $d$ 
is the
dimension of a spacelike hypersurface in Minkowski space. This is, of course,
very natural in case that the space of classical fields is a vector space as
well (before imposing the equations of motion), that is, for the case of 
scalar fields. That the quantum configuration space is required 
to be ${\cal S}'$ is furthermore justified 
1) by the fact that the Gausian measure underlying, say, the 
Klein-Gordon field is actually supported on ${\cal S}'$ and 2) by the 
Bochner-Minlos theory \cite{Yamasaki} which tells us that under natural
asuumptions about the characteristic functional of any measure (satisfied by
free scalar field theories) the support is guaranteed to be contained in
${\cal S}'$.

Secondly, the Euclidean group of spacetime plays an important role in that
the measure is required to be invariant under the motions of the Euclidean
group and that the time translation subgroup acts ergodically on the 
measure space. In particular, the discrete time reflection subgroup 
enables us to state the condition of ``reflection positivity" \cite{Jaffe}
by means of which one makes contact with physics, that is, one constructs 
a Hamiltonian and a Hilbert space. Again, all those properties are 
satisfied by the free scalar field theories and even some interacting
scalar field theories such as for instance the $P(\phi)_2$ theories.

However, what if the kinematical structure stated above is absent ?
We are particularly interested in a diffeomorphism invariant field
theory on a manifold with a fixed topology and whose classical configuration 
space is possibly non-linear. Then there is no Euclidean group but there 
is the diffeomorphism group and the measure of such a field theory should 
therefore be diffeomorphism invariant, rather than Euclidean invariant.
In general the Euclidean group will not even be a subgroup of the 
diffeomorphism group in question. 

Diffeomorphism invariant field theories are not at all of academic interest
only, rather, they are the only physically interesting field theories !
This is because the physical theory of nature is not the standard model 
but the standard model (or beyond) coupled to the {\em fully dynamical} 
gravitational field which is the theory of quantum gravity. This
field theory is indeed four-diffeomorphism invariant and contains 
kinematically non-linear fields such as gauge fields, Higgs fields and the
gravitational field.

Since a diffeomorphism invariant quantum field theory lacks the whole
arsenal of techniques that come with the presence of a fixed, 
non-dynamical Minkowski metric (the Poincar\'e group as symmetry group fixes 
vacuum and Hamiltonian operator, determines the irreducible 
representations under which the fields transform, the whole Wightman 
axiom system, a notion of time and so forth) we must expect that its
underlying kinematical structure is radically different from what one 
is used to from Euclidean field theory (whose Wick rotation, another 
process that depends on the presence of a Minkowski background metric,
results in the Wightman field theory). In particular, we do not expect the 
measure to be supported on ${\cal S}'$ any longer.

An already well-established example for such a departure from ${\cal S}'$ 
is provided by diffeomorphism invariant theories of 
connections for a compact gauge group in $d+1$ dimensions. In \cite{4}
such theories were canonically quantized and the representation of the 
canonical commutation relations that was chosen is generated by the Abelian 
$C^\star$ algebra $W$ of Wilson loop observables via the GNS construction 
\cite{0}. The quantum configuration space of distributional gauge fields
$\Ab$ (or $\agb$ after moding by the gauge freedom) is naturally 
identified with the compact Hausdorff space that one gets from $W$
by the Gel'fand isomorphism (recall that an Abelian $C^\star$ algebra is 
isometric isomorphic with the algebra of continuous functions on a compact
Hausdorff space). The Hilbert space of this class of theories, in other words
the measure on $\Ab$, is then uniquely determined by promoting the
reality structure of the classical phase space into adjointness 
relations with respect to the inner product. 

What is interesting from the point of view of constructive field theory
whose Hilbert space is determined by a measure on ${\cal S}'(\Rl^d)$ is 
that the space $\Ab$ is 1) not a vector space, 2) not a subset of
of ${\cal S}'(\Rl^d)$ and 3) ${\cal S}'(\Rl^d)$ is not contained in $\Ab$.
Thus, as argued above, taking the non-linear structure of the quantum 
configuration space seriously we obtain a fairly non-standard Hilbert space
in any spatial dimension $d$ which still represents the canonical 
commutation relations faithfully. Of course, from the point of view of,
say, the {\em free} Maxwell field (scattering processes) this Hilbert 
space is worthless (except 
in $d=2$ \cite{allmt1,almmt2}) because it does not coincide with Fock space.
However, it turns out as shown in \cite{4} that for a {\em diffeomorphism
invariant theory} the Hilbert space constructed is very useful : 
all the solutions of the diffeomorphism constraint are bona fide 
distributions on a dense space of test functions and therefore standard 
procedures known as ``group averaging techniques" \cite{Higuchi,Marolf}
lead to a well-defined inner product on this space of distributional
solutions \cite{4,6a}. The fact that for $d=2$ also Yang-Mills theory
is represented on this Hilbert space \cite{almmt2} can be traced back
to the fact that this theory is almost diffeomorphism invariant, it 
is invariant under area-preserving diffeomorphisms.\\
\\
In this paper we extend the canonical framework of \cite{4}, which 
applies to any kind
of gauge field theory, to the rest of the matter fields, specifically, 
we are looking at fermionic and Higgs field theories (including 
standard scalar fields which do not transform under any kind of gauge group
besides the diffeomorphism group).
Notice that \cite{4} also applies to quantum gravity when formulated as a 
dynamical theory of connections \cite{AA}. We need it in its manifestly 
real formulation as advocated, for the Lorentzian signature, first in 
\cite{Barbero} because \cite{4} only applies when the gauge group is compact
(in this case $SU(2)$).

We develop the framework for the Higgs field, in the 
diffeomorphism invariant context, by pushing the analogy between a usual 
path holonomy and a ``point holonomy" constructed from a Higgs Field (see
below) and can therefore translate step by step the framework 
\cite{0,1,2,3,3a} developed for gauge fields to arrive at the Higgs
field Hilbert space.

For the Fermion Field two works are to be mentioned :
in \cite{RM-T} the authors 
introduce an algebra of observables for complex Ashtekar gravity and the 
Maxwell field coupled to fermions. The fermions couple by making insertions
at the endpoints of Wilson lines. The authors compute the Poisson and 
commutator algebra of this (overcomplete) set of gauge invariant observables.
The framework was incomplete because no inner product for 
complex Ashtekar gravity was known at that time. Recently, the 
viewpoint for the kinematical framework of quantum gravity has shifted
towards a manifestly real formulation as advocated in \cite{Barbero} :
This is because 1) the rigorous kinematical framework of \cite{0,1,2,3,3a,4} 
can be employed and 2) the Wheeler-DeWitt constraint
operator can be completely rigorously formulated while with
complex variables \cite{5,6,6a} this turned out to be too difficult. 
Moreover, the same techniques can be 
employed to quantize a) Euclidean 2+1 gravity in the real-valued Ashtekar-
(rather than Witten-) formulation \cite{6b}
giving the expected results and b) the ADM energy \cite{6c}.\\
The authors of \cite{BK} have followed this trend and formulated a 
kinematical framework for diffeomorphism invariant theories of fermions
coupling to {\em real-valued} Ashtekar gravity and other gauge fields. 
That is, they define a natural inner product by means of the Berezin 
integral, show that a certain class of the operators defined in 
\cite{RM-T} are well-defined on it and indicate how to define 
diffeomorphism invariant distributions involving fermions along the lines 
of \cite{4}. However, certain difficulties having to do with the awkward
reality structure of the fermion field make their framework incomplete in 
the sense that most of the interesting operators defined in \cite{RM-T} 
do not have an adjoint on the Hilbert space defined. In particular,
this prevented the authors from proving that 
their inner product is selected by the classical reality conditions.
Since the adjointness relations on an inner product are the only conditions
that we can impose in lack of a background structure (in constructive 
field theory it is an {\em axiom} that the inner product be Poincar\'e 
invariant
which usually fixes it uniquely) it is unclear from their framework whether
their inner product is the appropriate one.

In this paper we define new fermionic canonical variables by casting the 
fermion field into a {\em half-density}. Not only does this trivialize
the classical reality conditions on the fermion field, it is also forced
on us : \\
If one does not work with half densities then it turns out that \\
a) The gravitational connection becomes complex valued. Thus, the
gravitational gauge group would become become $SL(2,\Co)$ which is 
non-compact and would make 
the arsenal of techniques developed in \cite{0,1,2,3,3a,4} inapplicable.\\
b) The faithful implementation of the reality conditions on scalar
fermion fields at the quantum level
is inconsistent with the implementation of the canonical Anti-Poisson
brackets as observed in \cite{BK}.\\ 
With the fermionic half-densities on the other hand 
we can then basically use the natural Berezin integral also employed in 
\cite{BK} to define the inner product. However,
now we can rigorously compute adjoints for any, not even gauge-invariant
operator constructed from fermion variables and then prove that the inner 
product is uniquely selected by the reality conditions. We do not need 
to deal with the complicated path observables defined in \cite{RM-T,BK}
in order to establish this result because we can prove everything at the 
non-gauge-invariant level.

The kinematical framework developed in this paper is applied in \cite{6d}
which extends the dynamical framework of \cite{5,6,6a,6b,6c} to the standard 
model coupled to gravity and other kind of matter.\\

The plan of the paper is as follows :\\
\\
In section 2 we introduce the notation and 
review \cite{0,1,2,3,3a,4}. 
Since we are using the canonical formalism, we assume that the four-dimensional 
manifold is locally of the form $M=\Rl\times\Sigma$ where $\Sigma$ can be any
three-dimensional manifold. The classical configuration subspace of  
the phase space can then be coordinatized by functions on $\Sigma$ which 
take values, respectively, in \\
a) the space of $SU(2)\times G$ connections $\omega_a$ for the gravitational 
and gauge field where $G$ is a compact gauge group,\\
b) the space of $Lie(G)$-valued scalars $\phi$  for the Higgs field and\\
c) the space of Grassman-valued fields $\eta$ transforming according to the 
fundamental representation of $SU(2)\times G$ for the Fermion Field.\\
We recall the refined algebraic quantization programme \cite{4}
whose major input is the choice of a $^\star$ representation of the classical
Poisson algebra.

In sections 3 and 4 we will then make the following choice of a $^\star$
representation :\\
We require that the following classical configuration observables 
(together with certain conjugate momentum observables) become densely 
defined operators on the Hilbert space : 
\begin{itemize}
\item[1)] Gravitational and Gauge Fields=Path Holonomies :\\
Given an open path $e$ in $\Sigma$ compute the holonomy (or path ordered 
exponential) $H_e(\omega)$, along $e$.
\item[2)] Higgs Field=Point Holonomy :\\
Given a point $p$ in $\Sigma$ and we may exponentiate the element of $Lie(G)$
corresponding to the Higgs field $\phi$ at $p$ to obtain a $G$-valued 
``point-holonomy" $U(p)$ (called this way due to its transformation 
properties under gauge transformations). 
\item[3)] Fermion Field=Grassman-valued Half-Density :\\
Given a point $p$ in $\Sigma$ we may evaluate the fermion field $\eta$ at 
this point to obtain a Grassman number $\eta(p)$ which transforms as a scalar
under diffeomorphisms. From this we construct the {\em half-density}
$\xi(p):=\root 4 \of{\det(q)}(p)\psi(p)$ where $q_{ab}$ is the 
dynamical metric on $\Sigma$. 
\end{itemize}
This choice is radical because the fields are not smeared with test 
functions in three space directions. In fact, 
the phase of the theory that we obtain will be very different from the 
Fock phase, a phase appropriate for a manifestly diffeomorphism invariant
description of the theory !\\
We show that the faithful implementation of the $^\star$ relations fixes 
the Hilbert space completely.

In section 5 we will finally combine 1),2),3) and obtain the 
gauge and diffeomorphism invariant Hilbert space by means of standard 
techniques \cite{Higuchi,Marolf}.

\section{Preliminaries}

We begin by describing the field content of the theory.

The topology of the four-dimensional manifold is chosen, as always
in the canonical approach, to be $M=\Rl\times\Sigma$ where $\Sigma$
is a smooth 3-manifold which admits smooth Riemannian metrics.  

The gravitational sector can be described \cite{Barbero} in terms of 
``real-valued Ashtekar variables", that is, by a canonical pair
$(A_a^i,E^a_i/\kappa)$ where $a,b,c$ etc. are tensorial indices for $\Sigma$,
$i,j,k$ etc. are $su(2)$ indices, $A_a^i$ is an $SU(2)$ connection 
and $E^a_i$ is its conjugate momentum, an $ad_{SU(2)}$ transforming vector 
density of weight one. The relation with the usual geometrodynamical 
variables is established by $E^a_i=\epsilon^{abc}\epsilon_{ijk}e_b^j e_c^k/2$
where $e_a^i$ is the co-triad and $q_{ab}=e_a^i e_b^i$ is the intrinsic 
metric of
$\Sigma$ while $A_a^i-\Gamma_a^i=\mbox{sgn}(\det((e_c^j)))K_{ab} e^b_i$
where $K_{ab}$ is the extrinsic curvature of $\Sigma$, $\Gamma_a^i$ the 
spin-connection of $e_a^i$ and $e^a_i$ its inverse. $\kappa$ is the 
gravitational constant.

Now we come to the matter sector. Let $G$ be an arbitrary compact gauge 
group, for instance
the gauge group of the standard model. Denote by $I,J,K,...$ $Lie(G)$
indices. We introduce classical Grassman-valued spinor fields 
$\eta=(\eta_{A,\mu})$
where $A,B,C,..$ denote indices associated with the gravitational $SU(2)$
and $\mu,\nu,\rho,..$ with the group $G$. Clearly, the fermion species 
$\eta$ transforms like a
scalar and according to an irreducible representation of $SU(2)\times G$.
It turns out that in its manifestly real form (the associated conjugation
is just complex conjugation for non-spinorial variables and for spinorial
fields it involves a cyclic reversal of order in products) the most 
convenient description of the constraints is in terms of half-densities
$\xi:=\sqrt[4]{\det(q)} \eta$. The momentum conjugate to $\xi_{A,\mu}$ is
then just given by $\pi_{A,\mu}=i\overline{\xi}_{A,\mu}$ :

Namely, as we will show in the next section, the gravitational connection of
the Einstein-Dirac theory is real valued (that is, it is an $su(2)$ 
rather than $sl(2,\Co)$ connection) only if one uses half-densities.
This is important because for complex valued connections the
techniques developed in \cite{1,2,3,3a,4} are inaccessable.\\
Notice that it is no lack of generality to restrict ourselves to just one
helicity as we can always perform the canonical transformation 
$(i\bar{\xi},\xi)\to(i\epsilon\xi,\overline{\epsilon\xi})$ where 
$\epsilon$ is the spinor-metric, the totally skew symbol in two 
dimensions. Notice that there is no  
minus sign missing in this canonical transformation because we take the 
fermion fields to be anti-commuting,
the action is invariant under this transformation \cite{Jacobson,6d}.

In the gauge sector we have canonical pairs 
$(\underline{A}_a^I,\underline{E}^a_I/Q^2)$ where the first entry is a $G$
connection and the second entry is the associated electric field, $Q$ is the
Yang-Mills coupling constant.
Finally, we may have scalar Higgs fields described by a canonical pair 
$(\phi_I,p^I)$ transforming according to the adjoint representation of $G$. 
Without loss of generality we can take these as real valued by suitably 
raising the number of Higgs families. Here and 
in what follows we assume that indices $I,J,K,..$ are raised and lowered 
with the Cartan-Killing metric $\delta_{IJ}$ of $G$ which we take to be 
semi-simple up to factors of $U(1)$.

Rarity-Schwinger fields can be treated by similar 
methods as here but we refrain from doing so because it is a simple 
exercise to apply the methods developed in this paper to supersymmetric 
field theories.

We will denote by $\tau_i$ the generators of the Lie algebra of 
$su(2)$ with the convention $[\tau_i,\tau_j]=\epsilon_{ijk}\tau_k$,
by $\omega_a:=A_a+\underline{A}_a$ the $SU(2)\times G$
connection. We have written only one family member of the possibly arbitrary
large family of field species, in particular, we can have an arbitrary number
of gauge fields all associated with different gauge groups and associated 
``quarks" and ``Higgs" fields transforming under different irreducible 
representations. Also, one could easily deal with a 
more complicated ``unified gauge group" which is not of the product type
$SU(2)\times G$ but contains it as a subgroup. However, for simplicity 
we will not deal with these straightforward generalizations and consider
only one field species transforming under the fundamental representation
of both $SU(2)$ and $G$ (fermions) or $G$ (Higgs field) respectively.
This furnishes the classical description of the field content.\\
\\
We now come to the quantum theory. We can immediately apply the techniques
of \cite{1,2,3,3a,4} to write down a kinematical inner product for the
gravitational and Yang-Mills sector that faithfully incorporates all the 
reality conditions. We get a Hilbert space 
$L_2(\Ab_{SU(2)}\times\Ab_G,d\mu_{AL,SU(2)}\otimes d\mu_{AL,G})$ where the 
index ``AL" stands for Ashtekar-Lewandowski measure and the group index
indicates to which gauge group the Ashtekar-Lewandowski measure is 
assigned. The reader interested
in the constructions and techniques around the space of generalized
connections modulo gauge transformations is urged to consult the papers
listed. In particular, the probability measure $\mu_{AL}$ is very natural and 
gauge and diffeomorphism invariant. 

The extension of the framework to Higgs and fermionic fields is 
the subject of this paper to which we turn in the next two sections.
The quantization programme consists, roughly, of the following steps (see 
\cite{4}) :\\
1) Choose a representation of the canonical commutation relations, that is, 
choose a vector space of functions $\Phi$ and an (over)complete set of basic 
operators corresponding to classical observables which leave $\Phi$ 
invariant such that the Poisson bracket relations among those functions 
become canonical (anti)commutation relations between the corresponding 
operators.\\ 
2) Choose an inner product $<.,.>$ on $\Phi$ such that the classical reality 
conditions among the basic functions become adjointness relations among 
the corresponding operators. Show that $<.,.>$ is unique, given the 
representation chosen, when requiring that the Poisson $^\star$ algebra is 
faithfully represented.\\
3) Complete $\Phi$ with respect to $<.,.>$ to obtain a kinematical
Hilbert space ${\cal H}$. The basic operators thus become densely defined.
Identify the gauge invariant subspace of $\cal H$.\\
4) Equip $\Phi$ with a topology of its own to make it a topological 
vector space. Apply the group averaging procedure of \cite{4} to construct
solutions to the diffeomorphism constraint which lie in the 
topological dual $\Phi'$ of $\Phi$ and a scalar product $<.,.>_{Diff}$ on 
these diffeomorphism invariant states.\\
5) So far all steps could be applied to all fields (gauge, fermion, 
Higgs) separately. Now put everything together and construct the 
gauge and diffeomorphism invariant states depending on all three kinds
of fields.

\section{Grassman-valued half-densities}

In this section we develop the integration theory for Grassman-valued 
half-densities since to our knowledge this is a situation usually not 
considered in the literature.

The reason for dealing with half-density-valued rather than scalar-valued
Grassman variables are two-fold :\\
\\
The first reason comes from the fact that in the latter 
case the momentum
conjugate to $\xi$ would be the density $\pi=i\sqrt{\det(q)}\overline{\xi}$
\cite{6d}.
The density weight itself is not troublesome, what is troublesome is that
the density weight comes from the function $\sqrt{\det(q)}$ which in quantum
gravity becomes itself an operator. This immediately leads to inconsistencies
as already stated in \cite{BK}
if one wants to construct an inner product on functions of the fermion fields
which incorporates the reality condition 
$\overline{\pi}=-i\sqrt{\det(q)}\xi$. Let us repeat the argument of
\cite{BK} for the sake of clarity.
Namely, assume, for instance, that we have managed to construct an 
inner product such that 
$\hat{\pi}^\dagger=-i\widehat{\sqrt{\det(q)}} \hat{\xi}$ and if $f=f(A)$ 
is any non-trivial real-valued function of the gravitational connection $A$ 
then $0=0^\dagger=([\hat{\pi},f(A)])^\dagger=
i[\widehat{\sqrt{\det(q)}},f(A)] \hat{\eta}\not=0$ which is a contradiction
unless the commutator vanishes which would be unnatural because the classical
Poisson bracket is non-zero.
Thus, a different approach is needed. In \cite{BK} the authors partly solve 
the problem by restricting the adjoint operation to a certain sub-algebra of 
``real elements" of the (gauge-invariant) observable algebra. However, as 
one can show, 
the action of the Hamiltonian constraint derived in section 3 does not 
preserve any cyclic subspace generated by that subalgebra. Moreover, we
will show now that if one does not work with half-densities, then the 
gravitational connection becomes actually complex-valued so that one 
could not apply the techniques associated with $\agb$ which hold only for 
compact gauge groups. \\

The second reason becomes evident when recalling that
for the Einstein-Dirac theory, formulated in terms of the variables 
$(A_a^i,E^a_i)$ \cite{6d} which have proved to be successful for 
quantizing the Wheeler-DeWitt constraint \cite{5,6}, the symplectic 
structure for each of the 
fermionic degrees of freedom is given by $\Theta=\frac{i}{2}\int_\Sigma 
d^3x\sqrt{\det(q)}[\overline{\eta}\dot{\eta}-\dot{\overline{\eta}}\eta]$
with scalar-valued $\eta$. Assume now that we would define 
$\pi:=i\sqrt{\det(q)}\overline{\eta}$ to be the momentum conjugate to $\eta$.
Upon an integration by parts we see that the symplectic structure becomes
up to a total time derivative equal to
$\int_\Sigma d^3x[\pi\dot{\eta}+\frac{1}{4}\pi\eta E_a^i\dot{E}^a_i]$
where $E_a^i$ denotes the inverse of $E^a_i$. But since $E^a_i$ is the 
momentum conjugate to $A_a^i$ this implies that the 
real-valued gravitational connection $A_a^i$ acquires a purely imaginary
correction term $-\frac{1}{4}\pi\psi E_a^i
=-\frac{i}{4} \bar{\eta}\eta e_a^i$ which leads to new 
mathematical difficulties because the techniques developed in
\cite{0,1,2,3,3a,4} break down for complex valued connections.
In fact, both problems are related because if one uses $\psi$ as
a canonical variable then $A$ is not real-valued as we just proved and 
then 
$0=[\hat{\pi},f(A)]^\dagger
=i[\widehat{\sqrt{\det(q)}}\hat{\eta},f(A)^\dagger]$
is not a contradiction but very messy and it will be hard to write 
$f(A)^\dagger$ for general $f$. In other words, in \cite{BK} the
contradiction arises from dealing with a non-real-valued 
(even when neglecting a total time drivative) symplectic 
structure which is physically inacceptable (``the world is real-valued").\\
We therefore follow the approach oulined below.\\
\\
The key observation is that if we introduce the Grassman-valued  
half-density 
\be \label{b1} 
\xi:=\root 4\of{\det(q)}\eta
\ee
then we have the identity $\Theta=\frac{i}{2}\int_\Sigma d^3x [\overline{\xi}
\dot{\xi}-\dot{\overline{\xi}}\xi]$ because the time derivative of the 
half-density factor drops out in the antisymmetrized sum. The last 
equation, upon dropping a total time 
derivative, can also be written as $\Theta=\int_\Sigma d^3x [i\overline{\xi}]
\dot{\xi}$ thus displaying
\be \label{b2}
\pi:=i\overline{\xi}
\ee
as the momentum conjugate to $\xi$ {\em without} that $A_a^i$ gets a 
correction thus leaving the compactness of the gravitational gauge group 
intact. Moreover, the reality condition is now very simple namely
\be \label{b3}
\overline{\pi}=-i\xi
\ee
which does not lead to the inconsistencies displayed above because the 
$\sqrt{\det(q)}$ does not show up in (\ref{b3}).

We begin with only one Grassman variable and recall some facts from 
Berezin integration \cite{Berezin}. Later we will generalize to the field 
theory case. All what we say here can be reformulated in terms of
supermanifold theory \cite{DeWitt} but since we only need a tiny
fraction of the apparatus we refrain from introducing the 
relevant notions and keep the vocabulary to a minimum.\\

Let $\theta$ be a Grassman variable and $\bar{\theta}$ its 
adjoint, enjoying the familiar anti-commutation relations 
$\theta^2=\bar{\theta}^2=\theta\bar{\theta}+\bar{\theta}\theta=0$ which 
state that $\theta,\bar{\theta}$ are Grassman-valued. On the other
hand, $\theta,\bar{\theta}$ obey an anti-Poisson algebra
$\{\theta,\theta\}_+=\{\bar{\theta},\bar{\theta}\}_+=0,\;
\{\theta,\bar{\theta}\}_+=-i$. Here $\{f,g\}_+=(-1)^{n_f n_g+1}\{g,f\}_+$
where $n_f\in\{0,1\}$ denotes the Grassman parity of $f$.

Any ``holomorphic" function of $\theta$ is of the form
$f(\theta)=a+b\theta$ where for our purposes it will be sufficient to take
$a,b$ to be complex numbers. A general function of $\theta,\bar{\theta}$ 
is of the form $F(\bar{\theta},\theta)=a+b\theta+c\bar{\theta}+d\bar{\theta}
\theta$. Then the ``measure" $d\bar{\theta}d\theta$ on the Grassman space
(or superspace) $\cal S$ coordinatized by $\theta,\bar{\theta}$
is defined by $\int_{\cal S} d\bar{\theta} d\theta F=d$. Superspace 
becomes a measurable space in the sense of measure theory \cite{Yamasaki}
upon equipping it with the trivial $\sigma$ algebra consisting only of
$\cal S$ itself and the empty set. This Berezin ``measure" is 
insufficient for our purposes because we want to construct an $L_2$ space
of holomorphic functions which requires a positive definite inner product.
We therefore define a positive measure on $\cal S$ 
by $dm(\bar{\theta},\theta):=e^{\bar{\theta}\theta}d\bar{\theta}d\theta=
(1+\bar{\theta}\theta)d\bar{\theta} d\theta$ and readily verify that
$<f,f>:=\int d\mu \bar{f} f=|a|^2+|b|^2$. Thus $||f||=0$ iff $f=0$ which 
would not be the case without the additional ``density" 
$dm/(d\bar{\theta}d\theta)$. Here we say that a function is 
positive if it is a linear combination of functions of the form $\bar{f}f$
where $f$ is holomorphic (i.e. depends only on $\theta$ but not on 
$\bar{\theta}$). Then the integral of positive functions is positive and 
therefore $dm$ is a positive measure.\\
As usual we now quantize the Grassman algebra by $\hat{\theta}f:=\theta f,\;
\widehat{\bar{\theta}}f:=df/d\theta$ which is easily checked to verify
$[\hat{\theta},\widehat{\bar{\theta}}]_+=1$ which shows that the anti 
Poisson algebra is faithfully represented provided that 
$\pi=i\bar{\theta}$ is the momentum conjugate to $\theta$ so that
$\hat{\pi}=id/d\theta$.\\
We must show that with respect to ${\cal H}:=L_2(dm,{\cal S})$ it holds that
$\hat{\theta}^\dagger=\widehat{\bar{\theta}}$. We compute
\ba \label{b4}
<f,\hat{\theta}g>&=&\int d\bar{\theta}d\theta (1+\bar{\theta}\theta)
\overline{f}\theta g\nonumber\\
&=&\int d\bar{\theta}d\theta \overline{f}\theta g\nonumber\\
&=&\int d\bar{\theta}d\theta \overline{[\theta \frac{df}{d\theta}]}\theta 
g\nonumber\\
&=&\int d\bar{\theta}d\theta \overline{\frac{df}{d\theta}}\bar{\theta}\theta 
g\nonumber\\
&=&\int d\bar{\theta}d\theta [\bar{\theta}\theta]
\overline{\frac{df}{d\theta}} g\nonumber\\
&=&\int d\bar{\theta}d\theta (1+\bar{\theta}\theta)
\overline{\frac{df}{d\theta}} g\nonumber\\
&=& <\widehat{\bar{\theta}}f,g>\;.
\ea
In the second equality we used that $\theta^2=0$ in order to get rid of 
the $\bar{\theta}\theta$, in the third we used that $d\bar{\theta} d\theta$ 
picks the term proportional to $\bar{\theta}\theta$ which comes from the 
term of $f$ linear in $\theta$ and the factor of $\theta$ already 
present,
in the fourth and fifth we used the definition of the involution of the 
Grassman variables and that $\bar{\theta}\theta$ is bosonic, in the sixth
we recoverd the density of the measure because the additional integral is 
identically zero since $\overline{df/d\theta}g$ cannot have a term 
proportional to $\bar{\theta}\theta$. Thus, we verified that the reality
conditions are faithfully implemented. \\
We wish to show now that, given the representation on holomorphic 
functions of the anti-Poisson bracket algebra defined by
$\hat{\theta}f= \theta f,\;\hat{\pi}f=i df/d\theta$, the measure 
$dm$ is the {\em unique} probability measure selected by asking that 
$\hat{\theta}^\dagger=\widehat{\bar{\theta}}$. This follows from considering 
the most general Ansatz for a 
measure given by $dm(\bar{\theta},\theta)=F(\bar{\theta},\theta)
d\bar{\theta}d\theta$ with 
$F=a+b\theta+c\bar{\theta}+d\bar{\theta}\theta$ as above and 
studying the implications of 
asking that $<1,1>=1,\;<f,\theta g>=<\frac{df}{d\theta},g>,\;
<f,\frac{dg}{d\theta}>=<\theta f,g>$. We leave it as anexercise to check 
that indeed the first condition leads to $d=1$, the second to $a=1,c=0$ and 
the third to $b=0$. Thus ${\cal H}:=L_2({\cal S},dm)$ is the unique
Hilbert space selected by the $^\star$ relations.

These results are readily generalized to the case of a finite number, say 
$n$, of Grassman variables with the anti-Poisson algebra 
$\{\theta^i,\theta^j\}_+
=\{\bar{\theta}^i,\bar{\theta}^j\}_+=0,\{\bar{\theta}^j,\theta^k\}_+=
-i\delta^{ij}$. Now a holomorphic function has the form 
$$f=\sum_{k=0}^n\frac{1}{k!} a_{i_1..i_k}\theta^{i_1}..\theta^{i_k}$$
with totally skew complex valued coefficients $a_{i_1..i_k}$. We 
introduce the product measure $dm(\{\bar{\theta}^i\},\{\theta^i\})
=\prod_{i=1}^n dm(\bar{\theta}^i,\theta^i)$ and find that
\be \label{b4a}
<f,f>=\sum_{k=0}^n\sum_{1\le i_1<..<i_k\le n}|a_{i_1..i_k}|^2
\ee
which shows that $L_2({\cal S},dm)$ is a $2^n$ dimensional Hilbert space with
orthonormal basis $\theta^{i_1}..\theta^{i_k},\;k=0,..,n,\; 1\le 
i_1<..<i_k\le n$. Moreover, by defining $\hat{\theta}^i f=\theta^i 
f,\;\widehat{\bar{\theta}}^i f =\partial f/\partial\theta^i$
the anti-Poisson algebra is faithfully represented together with the $^*$
relations.\\ 
This furnishes the quantum mechanical case.

Let us now address the field theoretic case. Recall that we have the 
classical anti-Poisson brackets $\{\xi_{A\mu}(x),\pi_{B\nu}(y)\}_+
=\delta_{AB}\delta_{\mu\nu}\delta(x,y)$. Notice that on both sides of 
this equation we have a density of weight one. Moreover, we have the 
classical $^*$ relations 
$\overline{\pi_{A\mu}}=\pi_{A\mu}^\star=-i\xi_{A\mu}$.
Let now $f$ be a general holomorphic functional of the $\xi_{A\mu}(x)$.
We define operators $\hat{\xi}_{A\mu}(x)f:=\xi_{A\mu}(x)f,\;
\hat{\pi}_{A\mu}(x)f:=i\hbar\frac{\delta f}{\delta\xi_{A\mu}(x)}$ or, 
eqivalently,
$\widehat{\bar{\xi}}_{A\mu}(x)f:=\hbar\frac{\delta f}{\delta\xi_{A\mu}(x)}$.
It is easy to see that this satisfies the required anti-commutator 
relation 
$[\hat{\xi}_{A\mu}(x),\hat{\pi}_{B\nu}(y)]_+=i\hbar\delta_{AB}
\delta_{\mu\nu}\delta(x,y)$ by using that the functional derivative is also
Grassman-valued. \\
\\
Now, however, we must take seriously into account that $\xi(x)$ is a 
half-density (we will suppress the label $A\mu$ in this paragraph). 
The first observation is that it poses no problem to 
extend the above framework to the case of countably infinite number of 
variables. Let us therefore introduce a triangulation 
$\Sigma=\cup_n B_n$ of $\Sigma$, that is, a splitting into 
countably many, mutually disjoint (up to common faces, edges and 
vertices) closed solid boxes $B_n$ such 
that $\int_\Sigma=\sum_n\int_{B_n}$. Each $B_n$ has a Lebesgue
measure $\epsilon_n^3$ and a centre $v_n=v(B_n)$. Let $\chi_n(x):=
\chi_{B_n}(x)$ denote the characteristic function of $B_n$.
Let us define new Grassman-valued variables 
\be \label{b100}
\theta_n:=\int_\Sigma \frac{\chi_n(x)}{\sqrt{\epsilon_n^3}} \xi(x)\;.
\ee
Then it is easy to see that 
\be \label{b101}
\sum_n \bar{\theta}_n\theta_n=\sum_n\int d^3x \chi_n(x) 
\int d^3y\frac{\chi_n(y)}{\epsilon_n^3} \bar{\xi}(x)\xi(x)
\to \int d^3x \bar{\xi}(x)\xi(x) \mbox{ as }\epsilon_n\to 0\forall\;n\;.
\ee
Furthermore, the corresponding operators 
\ba \label{b102}
\hat{\theta}_n&:=&\int_\Sigma \frac{\chi_n(x)}{\sqrt{\epsilon_n^3}} 
\hat{\xi}(x),\nonumber\\
\widehat{\bar{\theta}}_n&:=&\int_\Sigma 
\frac{\chi_n(x)}{\sqrt{\epsilon_n^3}} \widehat{\bar{\xi}}(x)
\ea
enjoy discrete canonical anti-commutator relations (even at finite 
$\epsilon_n$)
\ba \label{b103}
 [\hat{\theta}_m,\hat{\theta}_n]_+ 
&=& [ \widehat{\bar{\theta}}_m,\widehat{\bar{\theta}}_n ]_+=0 
\nonumber\\
{[} \widehat{\bar{\theta}}_m,\hat{\theta}_n {]}_{+} 
&=& \int d^3x\int d^3y 
\frac{\chi_m(x)\chi_n(y)}{\sqrt{\epsilon_m^3\epsilon_n^3}}
 [\widehat{\bar{\xi}}(x),\hat{\xi}(y)]_+
\nonumber\\
&=& \int d^3x\frac{\chi_m(x)\chi_n(x)}{\sqrt{\epsilon_m^3\epsilon_n^3}}
=\delta_{m,n}\int d^3x\frac{\chi_m(x)}{\epsilon_m^3}=\delta_{m,n}
\ea
because $\chi_m(x)\chi_n(x)=\delta_{m,n}\chi_n(x)$. A similar calculation
using the anti-bracket relation $\{\xi(x),\bar{\xi}(y)\}_+=-i\delta(x,y)$
reveals that $\{\theta_m,\bar{\theta}_n\}=-i\delta_{m,n}$.\\

The quantities $\theta_n$ have a {\em formal} limit as $\epsilon_n\to 0$
(i.e. the triangulation becomes the continuum) and $B_n\to 
v(B_n)=:x$ : Namely, because $\chi_n(x)=\sqrt{\chi_n(x)}$ it follows 
that 
\be \label{b104}
\theta(x):=\lim_{\epsilon_n\to 0}\theta_n=\lim_{\epsilon_n\to 0}\int 
d^3y\sqrt{\frac{\chi_n(y)}{\epsilon_n^3}}\xi(y)
``="\int_\Sigma d^3y\sqrt{\delta(x,y)}\xi(y)
\ee
where $\delta(x,y)$ is the three-dimensional $\delta$-distribution.
Notice that the $\delta$ distribution is a density of weight one
so that due to the square root and the half-density $\xi$ the quantities 
$\theta(x)$ {\em transform as scalars} ! This is important for the 
construction of diffeomorphism invariant states as we will see below.\\

We are now ready to construct an inner product. At finite $\epsilon_n$ we 
define a regulated Fermion measure by
\be \label{b105}
d\mu_F^\epsilon(\bar{\theta},\theta):=\prod_n dm(\bar{\theta}_n,\theta_n)
\ee
where $\epsilon:=\inf_n\{\epsilon_n\}>0$ and $dm$ is the measure consructed 
above. It is easy to see that the $^\star$ relations 
$\widehat{\bar{\theta}}^\dagger_n=\hat{\theta}_n$ hold for each $n$.
The formal limit, as 
$\epsilon\to 0$, of this measure is given by 
\ba \label{b106}
d\mu_F(\bar{\xi},\xi)&:=&
(\prod_{x\in\Sigma} [d\bar{\theta}(x) d\theta(x)])
e^{\sum_{x\in\Sigma} \bar{\theta}(x)\theta(x)}
\nonumber\\
&=&(\prod_{x\in\Sigma} [d\bar{\theta}(x) d\theta(x)])
e^{\int_\Sigma d^3x \bar{\xi}(x)\xi(x)}
\ea
In the first step we have used the fact that the quantities 
$\bar{\theta}_n\theta_n$ mutually commute to write the product of the 
exponentials as the exponential of the sum of exponents and then took the 
limit. In the second step we exploited (\ref{b101}) in addition.
By inspection, the measure $d\mu_F$ is diffeomorphism invariant and obviously
one formulates the theory better in terms of the $\theta(x)$ than in terms
of the half-densities $\xi(x)$.\\ 
All identities that we will state in the 
sequel can be rigorously justified by checking them at finite $\epsilon$ 
and then take the limit. For example, we already verified that
$\widehat{\bar{\theta}}^\dagger_n=\hat{\theta}_n$ for all $n$ with 
respect to $\mu_F^\epsilon$. Taking $\epsilon\to 0$ we conclude that 
\be \label{b107}
\widehat{\bar{\theta}}^\dagger(x)=\hat{\theta}_n(x)
\ee
for all $x$ with respect to $\mu_F$. This is obviously equivalent to 
the statement 
$\widehat{\bar{\xi}}^\dagger(x)=\hat{\xi}(x)$ for all $x$ with 
respect to $\mu_F$ in view of (\ref{b104}). Thus we have implemented the
reality conditions faithfully and the measure $\mu_F$ is the unique 
probability measure that does it (the uniqueness statement follows 1) from 
the uniqueness for each $m_x=m(\bar{\theta}(x),\theta(x))$ separately 
as seen above and 2) from the fact that all points of $\Sigma$ are 
uncorrelated).\\
Another example is the anti-commutator
$[\widehat{\bar{\theta}}_m,\hat{\theta}_n]_+=\hbar\delta_{m,n}$ whose 
limit is given by 
\be \label{b108}
[\widehat{\bar{\theta}}(x),\hat{\theta}(y)]_+=\hbar\delta_{x,y}\;.
\ee
Notice that $\delta_{x,y}$ is 
a Kronecker $\delta$ not a $\delta$-distribution which is precisely in 
accordance with the classical 
symplectic structure $\{\theta(x),\theta(y)\}=-i\delta_{x,y}$. In fact,
we have by the same reasoning as in (\ref{b101})
$$
\sum_n \bar{\theta}_n\dot{\theta}_n\to\sum_x \bar{\theta}_x\dot{\theta}_x
=\int d^3x \bar{\xi}(x)\dot{\xi}(x)
$$
as $\epsilon\to 0$.\\
This motivates the following.\\ 
\\
{\em We will choose a representation for which the 
scalar Grassman-valued field $\hat{\theta}(x)$ is the basic quantum 
configuration variable, in particular, it is densely defined.}\\ 
\\
As a last example we quote the following orthogonality relations 
$<\theta_m,\theta_n>=\delta_{m,n}$ which extend to 
$<\theta(x),\theta(y)>=\delta_{x,y}$.

Summarizing and re-introducing the labels $A\mu$, the Fermion measure is 
given by 
\be \label{b5}
d\mu_F([\bar{\theta}],[\theta]):=\prod_{x\in\Sigma} 
dm(\{\bar{\theta}_{A\mu}(x)\},\{\theta_{A\mu}(x)\})=:\prod_{x\in\Sigma} dm_x
\ee
where $dm_x$ is the measure on the Grassman space ${\cal S}_x$ at $x$ 
which is defined by the 
$2\times 2d$ Grassman variables at $x$
introduced above, $d$ the dimension of the fundamental representation of 
$G$. This measure is more precisely defined as follows : we say that a 
function $f$ is cylindrical whenever it depends only on the fermion fields 
at a 
finite number of points $v_1,..,v_n$ which in our case turn out to be the
vertices of a piecewise analytic graph $\gamma$. We then have the simple 
definition $\mu_F(f)=(m_{v_1}\otimes..\otimes 
m_{v_n})(f)=:m_{v_1,..,v_n}(f)$. 
These cylindrical product measures are consistently defined, that is, we have
$m_{v_1,..,v_n}(f)=m_{v_1,..,v_m}(f)$ whenever $f$ depends only on the 
fermion fields on the fermion fields at the first $m\le n$ points. Since
then we have an uncountable product of probability measures the Kolmogorov
theorem for the uncountable direct product \cite{Yamasaki} tells us that the 
consistent family $\{m_{v_1,..,v_n}\}$ has the unique $\sigma-$additive 
extension (\ref{b5}) to the infinite product measurable space
$\overline{{\cal S}}:=\times_x {\cal S}_x$. 
The Hilbert space is just given by 
$$
{\cal H}_F=L_2(\overline{{\cal S}},d\mu_F)=\otimes_x L_2({\cal S}_x,dm_x).
$$
Thus, ${\cal H}_F:=L_2({\cal S},d\mu_F)$ is the appropriate and unique 
kinematical
Hilbert space on holomorphic Grassman functionals (in terms of
$\theta(x)$) satisfying the appropriate
adjointness relations.
Notice that the measure $\mu_F$ is diffeomorphism invariant in the 
sense that it assigns the same integral to any cylindrical function before 
and after mapping the points $\vec{v}$ with a diffeomorphism. 
It is also gauge invariant since a gauge transformation at $x$ is in 
particular a unitary transformation on the on the $2d$ dimensional vector 
space with respect to the inner product (\ref{b4a}).

An orthonormal basis for $\mu_F$ can be constructed as follows :\\
Order the labels $A\mu$ in some way from $i=1$ to $i=2d$.
Let $\vec{v}$ be an ordered, finite set of mutually different points as 
above. For each $v\in\vec{v}$
denote by $I_v$ an array $(i_1(v)<..<i_k(v)$ where $0\le k=\le 2d$ and 
$1\le i_j(v)\le 2d$ for each $1\le j\le k$. Also, let $|I_v|=k$ in this 
case. Finally 
let $\vec{I}:=\{I_v\}_{v\in\vec{v}}$. \begin{Definition} \label{fermvert}
A fermionic vertex function $F_{\vec{v},\vec{I}}(\theta)$ is 
defined by $F_{\vec{v},\vec{I}}:=\prod_{v\in\vec{v}}F_{v,I_v}$
where $F_{v,I_v}:=\prod_{j=1}^k\theta_{i_j(v)}(v)$.
\end{Definition} 
It follows that a cylindrical function $f_{\vec{v}}$ can be written as
\be \label{b5a}
f_{\vec{v}}=\sum_{\vec{I}} a_{\vec{v},\vec{I}} F_{\vec{v},\vec{I}}
\ee
for some complex valued coefficients $a_{\vec{v},\vec{I}}$ and the sum runs
for each $v\in\vec{v}$ over the $2^{2d}$ values of $I_v$.
\begin{Lemma}
The collection of fermionic vertex functions provides an orthonormal 
basis for ${\cal H}_F$.
\end{Lemma}
Proof :\\
By construction the vertex functions provide a dense set of vectors
in ${\cal H}$. To see the orthonormality one just needs to recall
(\ref{b4a}).\\
$\Box$\\
\\
We now proceed to the solution of the diffeomorphism constraint. \\
First of all, let us denote by $\Phi_F$ the space of cylindrical functions
constructed only from fermionic fields. Thus, any element is a finite 
linear combination of products of the functions $F_{v,I}$ of 
Definition \ref{fermvert}. We equip 
$\Phi_F$ with a topology by assigning to the element $f_{\vec{v}}$ the 
``Fourier-norm" 
$$
||f_{\vec{v}}||_1:=\sum_{\vec{I}}|<F_{\vec{v},\vec{I}},f_{\vec{v}}>|
$$
the name being motivated by the fact that this is like an $L_1$ norm on 
the ``Fourier coefficients" $a_{\vec{v},\vec{I}}$ in (\ref{b5a}).
The space $\Phi'_F$ is the topological dual of $\Phi_F$ with respect
to $||.||_1$. We then have the inclusion $\Phi_F\subset{\cal H}_F\subset
\Phi'_F$ since $||f||_2^2\le ||f||_1^2$ so that the assertion follows from 
the Schwarz inequality.

We wish to construct distributions in 
$\Phi_F'$ which are diffeomorphism invariant with respect to 
smooth diffeomorphisms
(in case that the points 
$\vec{v}$ of a vertex function are actually vertices of a 
piecewise analytic $\gamma$ graph with vertex set $V(\gamma)$ defined by
the gauge fields, consider only analyticity 
preserving smooth diffeomorphisms \cite{6a}). 
This is straightforward given the framework of \cite{4} : 
Notice that a unitary representation of the diffeomorphism group on 
${\cal H}_F$ is defined by $\hat{U}(\varphi)f_{\vec{v}}:=
f_{\vec{\varphi(v)}}$. This transformation law under diffeomorphisms 
is compatible with the fact that the $\theta_i(x)$ are {\em scalars}
rather than half-desities. This is rather important because as we will see
later, there is no obvious way to group average the $\xi_i(x)$ such that the
result is a distribution.\\
Consider the orbit of the basis vector $F_{\vec{v},\vec{I}}$, that is,
consider the set of states 
$\{F_{\vec{v},\vec{I}}\}:=\{\hat{U}(\varphi)F_{\vec{v},\vec{I}};\;
\varphi\in \mbox{Diff}(\Sigma)\}$. We define the group average 
of the basis vectors $F_{\vec{v},\vec{I}}$ by
$$
[F_{\vec{v},\vec{I}}]:=\sum_{F\in\{F_{\vec{v},\vec{I}}\}} F\;.
$$
The group average of a general element (\ref{b5a}) is defined by 
requiring linearity of the group averaging procedure, that is,
$$
[f_{\vec{v}}]:=\sum_{\vec{I}}a_{\vec{v},\vec{I}}[F_{\vec{v},\vec{I}}]\;.
$$
Why the group averaging has to be done in terms of a chosen basis is 
explained in \cite{6a}. Since we are eventually interested in averaging
gauge invariant states the problem of the ``graph symmetry factors" outlined 
in \cite{6a} is resolved in the process of averaging the gauge fields.
Finally, the diffeomorphism invariant inner product is given by \cite{4,6a}
$$
<[f],[g]>_{Diff}:=[f](g)
$$
where the latter means evaluation of the distribution $[f]\in\Phi'_F$ on 
the cylindrical function $g\in \Phi_F$.\\
\\
We conclude this section with showing why the shift from $\xi(x)$ to
$\theta(x)$ becomes mandatory in the diffeomorphism invariant context :\\
Consider a cylindrical function $f_{v_1,..,v_n}$ with $n_v\le 2d$ 
fermionic half-densities $\xi$ at the point $v\in\{v_1,..,v_n\}$. Consider 
the transformation 
law of that function under a diffeomorphism $\varphi\in\mbox{Diff}(\Sigma)$.
Due to the half-density weight we get
$f_{v_1,..,v_n}\to \prod_{k=1}^n\frac{1}{J_\varphi(v)^{n_v}}
f_{\varphi(v_1),..,\varphi(v_n)}$ where 
$J_\varphi(v)=|\det(\partial\varphi^a/\partial x^b)(v)|^{1/2}$. Now let 
$\varphi$ be an (analyticity
preserving) diffeomorphism which leaves the points $v_1,..,v_n$ invariant
(and the underlying graph) but such that the $J_\varphi(v)\not=1$. Such
diffeomorphisms certainly exist as the reader can convince himself by
considering in two dimensions the diffeomorphism $\varphi(x,y)=(\alpha^2
x,y)$ which leaves, say, the interval $[0,1]$ along the $y$-axis 
invariant but $J_\varphi(0,0)=|\alpha|$.  We see that all the states 
$\lambda 
f_{v_1,..,v_n},\lambda>0$ are related by a diffeomorphism and therefore
the group average of this state along the lines of \cite{4} would include 
the object $[\sum_{\lambda>0} \lambda]f_{v_1,..,v_n}$ which is meaningless 
even when considered as a distribution on $\Phi_F$ because it has 
non-vanishing
inner product with $f_{v_1,..,v_n}\in \Phi_F$. Notice that 
for scalar valued fermions $\theta$ instead of $\xi$ on the other hand 
the diffeomorphisms just discussed would not alter the orbit of the state 
and therefore would not contribute to the group average.

\section{Point Holonomies}

We have made for the gravitational and gauge fields the (rather 
radical) assumption that the holonomies along edges can be promoted to a
densely defined operator. This assumption has lead to the Ashtekar-Isham
quantum configuration space $\agb$ of distributional connections modulo 
gauge transformations and to the Ashtekar-Lewandowski measure $\mu_{AL}$
which is the unique measure (up to a positive constant) that promotes the 
reality conditions on the classical phase space into adjointness 
relations on the corresponding Hilbert space $L_2(\agb,d\mu_{AL})$.

In this section we are going to make an even more radical assumption for 
the Higgs field : Namely, given a classical Higgs configuration field 
$\phi_I(v)$ we may exponentiate it to obtain the so-called {\em holonomy
at the point $v$}, $U(v):=\exp(\phi_I(v)\tau_I)$, where $\tau_I$ are the 
generators of $Lie(G)$. The $U(v)$ are then $G$-valued classical objects 
which transform under the adjoint representation of $G$ since
$\phi_I\tau_I\to g[\phi_I(v)\tau_I]g^{-1}$. Here and in what 
follows we consider only unitary
and unimodular compact groups (so that the Cartan Killing metric is given
by $\delta_{IJ}$). The case of a scalar field would correspond
to $U(v)=e^{i\phi(v)}\in\;U(1)$ and can be handeled similarily.
The key assumtion is now \\
\\
{\em The point holonomies $U(v)$ can be promoted to densely defined operators
on a Hilbert space that implements the appropriate adjointness relations !}\\
\\
The adjointness relations that we would like to implement are, of course,
that (in a sense to be made precise below) 
$[\phi_I(v)]^\dagger=\phi_I(v),\;[p^I(v)]^\dagger=p^I(v)$.\\
\\
Let us first pause and give a rational why it is more appropriate,
in the diffeomorphism invariant context, to base the quantization on the 
point holonomies $U(v)$ rather than on the Higgs (or scalar) fields 
$\phi_I(v)$ itself :\\
Assume that we choose as our basic configuration field variables the 
$\phi_I(x)$. Having made this assumption, then a natural
choice for the inner product for scalar fields is a Gaussian measure : some 
sort of Gaussian joint distribution is needed because $\phi_I$ 
is real valued rather than valued in a compact set. We will now argue 
that this sort of a kinematical measure is in conflict with 
diffeomorphism invariance when one implements the canonical commutation 
relations and the adjointness relations (we consider the case of scalar 
field $\phi$, even more problems associated with gauge invariance result 
with Higgs fields).
In an attempt to resolve the problems that occur we will see that 
we are naturally lead to the consideration of point holonomies.
  
Recall \cite{Jaffe} that a measure for a field whose quantum 
configuration space is a vector space is rigorously 
defined in terms of its characteristic functional. The characteristic 
functional is given by the integral of the functional
$e^{i\phi(f)}$, that is, $\chi(f):=\mu(e^{i\phi(f)})$ where 
$\phi(f)=\int_\Sigma d^3x f(x)\phi(x)$ and $f$ is some sort of real 
valued test function
on $\Sigma$ (usually of rapid decrease). The point is now that the label 
$f(x)$ is a function of the coordinates $x^a$ whose only 
meaningful transformation law under diffeomorphisms amounts to saying that 
they are scalars and therefore $f$ is a scalar. Therefore the pairing 
$\phi(f)$ only has a covariant meaning (observing that $\Sigma$ is 
invariant under diffeomorphisms) provided that $\phi$ is a scalar density !
This can be achieved by performing the classical canonical transformation
$(\phi,p)\to(p,-\phi)$ and so does not pose any immediate problem.
In this case we have a representation of the diffeomorphism group given
by $\hat{U}(\varphi)\phi(f)=\phi([\varphi^{-1}]^\ast f)$. Therefore,
the measure $\mu$ is diffeomorphism invariant if and only if
$\chi(\varphi^\star f)=\chi(f)$ for any $\varphi$. Now a Gaussian measure
is defined through its covariance $C$ which is the kernel
of an operator in the sense that 
$$
\chi(f)=e^{-\frac{1}{2}C(f,f)}\mbox{ where } C(f,g):=
\int d^3x\int d^3y f(x) C(x,y) g(y)\;.
$$
Obviously diffeomorphism invariance imposes that\\
$C(x,y)=|\det(\partial\varphi(x)/\partial x)
\det(\partial\varphi(y)/\partial y)|C(\varphi(x),\varphi(y))$
which is the transformation law of a {\em double} density. The only quantity
merely built from the coordinates, which transfroms like a double 
density and is diffeomorphism invariant is of the form
$C(x,y)=\sum_{z\in\Sigma} \delta(z,x)\delta(z,y)$ (notice that this 
involves a sum rather than an integral over $\Sigma$). But 
$\sum_{z\in\Sigma}|f(z)|^2$ does not 
diverge only if the test function $f$ is not smooth but vanishes at all 
but a finite number of points. But then $\phi(f)=0$ for almost
all tempered distributions $\phi$ except for those which are of the 
form $\phi(x)=\sum_{z\in\Sigma}\delta(x,z)\Phi(z)$ where $\Phi(z)$ is a 
scalar. We see that $\exp(i\phi(f))$ equals the point holonomy
$\exp(i\sum_{x\in\mbox{supp}(f)}f(x)\Phi(x))$ which therefore comes out 
of the analysis naturally. But let us further stick with the assumption
that the $\phi(x)$ can be made well-defined operators.
Then the diffeomorphism invariant measure automatically is supported on those 
$\phi$ only which are of the form $\sum_{z\in\Sigma}\delta(x,z)\Phi(z)$.
The corresponding Gaussian measure is formally given by
$[\frac{d\Phi}{2\pi}]\exp(-1/2\sum_{z\in\Sigma}\Phi(z)^2)$, a 
``discrete white noise measure" which is not
even expressible in terms of $\phi(x)$. We see that the measure is an 
infinite product of Gaussian measures, one for each point in $\Sigma$.\\
We see that the assumption that $\phi$ was a density has lead to a more
natural description in terms of scalar $\Phi$ so let us start with
scalar $\phi:=\Phi$ right from the beginning and 
that the Gaussian measure is the discrete white noise measure in terms of 
$\phi(x)$. Then multiplication by $\phi(x)$ and the operator 
$-i\hbar[\delta/\delta\phi(x)-1/2\sum_{z\in\Sigma}\delta(x,z)]=\hat{p}(x)$ 
smeared 
with a three-dimensional test function become formally self-adjoint
operators on the associated $L_2$ space which has an orthonormal basis 
given by finite linear combinations of 
Hermite monomials $:\phi(x_1)^{n_1}:..:\phi(x_m)^{n_m}:$ (Wick ordering 
with respect to the Gaussian measure above). The trouble with this 
approach is that multiplication by $\phi(x)$ but not  
$\hat{p}(x)$ are densely defined, not even when smeared with a 
test function : to see this consider the simplest state, the ``vacuum state "
$f(\phi)=1$. Let $B$ be a compact region in $\Sigma$, then
$f':=\int_B d^3y \hat{p}(y) f=i\hbar/2\sum_{z\in B}\phi(z)$ which is a state 
of infinite norm because $<\phi(x),\phi(y)>=\delta_{x,y}$, for any compact 
set $B$. Thus, we conclude that the Gaussian measure defined above and 
which comes from the assumption that the $\phi$ are well-defined operators
leads to rather difficult mathematical problems given our adjointness 
relations. \\ \\
The rest of this section is devoted to constructing an appropriate quantum
configuration space $\overline{{\cal U}}$ of Higgs fields where 
$\cal U$ denotes the space of classical, that is, smooth Higgs fields and 
to determine a unique $\sigma-$additive measure $\mu_U$ on 
$\overline{{\cal U}}$ selected by the adjointness relations. As we will see,
we can roughly copy the procedure of \cite{0,1}.
\begin{Definition} \label{defc1}
A Higgs vertex function is a function 
$H_{\vec{v},\vec{\pi},\vec{i},\vec{j}}$ on 
$\cal U$ labelled by a quadruple $(\vec{v},\vec{\pi},\vec{i},\vec{j})$.
Here $\vec{v}:=\{v_1,..,v_n\}$ is a finite set of points in $\Sigma$ and
$\vec{\pi}:=\{\pi_1,..,\pi_n\}$ is a corresponding set of irreducible 
representations of $G$ (choose once 
and for all a representant from the class of independent equivalent 
representations).
Finally, $\vec{i}:=\{i_1,..,i_n\},\;\vec{j}:=\{j_1,..,j_n\}$
are corresponding labels for the matrix elements of $\pi_1,..,\pi_n$.
We say that the point $v_i$ is dressed by $\pi_i$ and obtain a complex
valued function
\be \label{c1}
H_{\vec{v},\vec{\pi},\vec{i},\vec{j}}(\phi):=
[\otimes_{k=1}^n [\sqrt{d_{\pi_k}}\pi_k(U(v_k))]]_{\vec{i},\vec{j}}:=
\prod_{k=1}^n [\sqrt{d_{\pi_k}}\pi_k(U(v_k))]_{i_k j_k}
\ee
where $d_\pi$ denotes the dimension of the irreducible representation $\pi$.
\end{Definition}
Let us now consider the $^\star$ algebra $\tilde{V}$ (with unit $1$, the 
constant function with value one) of functions 
consisting of the finite linear combinations of vertex functions where 
the $^\star$ operation is just complex conjugation. We turn it into an 
Abelian $C^\star$ algebra $V$ by completing it with respect to the 
supremum norm,
$||f||:=\sup_{\phi\in{\cal U}}|f(\phi)|\forall f\in\tilde{V}$. Notice that
the supremum norm makes sense because $|\pi_{ij}(U(v))|\le 1$.
\begin{Definition}
The space $\overline{{\cal U}}$ of distributional Higgs fields is the 
Gel'fand spectrum of $V$ equipped with the Gel'fand topology.
\end{Definition}
Recall that the Gel'fand spectrum of an (Abelian) $C^\star$ algebra is the 
set of its characters, that is, the set of {\em all} $^\star$ homomorphisms 
from $V$ into the complex numbers.
That is, given an element $\phi\in{\cal U}$ and an element $f\in V$ we obtain
a complex number $\phi(f)$, moreover, 
$\phi(f+g)=\phi(f)+\phi(g),\;\phi(fg)=\phi(f)\phi(g),\;\phi(f^\star)=
\overline{\phi(f)},\;\phi(1)=1$. This shows that ${\cal 
U}\subset\overline{\cal 
U}$. The Gel'fand topology on the spectrum is the weak
$^\star$ topology, that is, a net $\phi_\alpha$ in ${\overline{\cal U}}$
converges if the net of complex numbers $\phi_\alpha(f)$ converges for 
each element $f\in V$ pointwise. \\
The Gel'fand isomorphism $V\to C({\cal U});\; f\to\hat{f};\;
\hat{f}(\phi):=\phi(f)$ shows that $V$ is isomorphic with the set of
continuous functions on its spectrum. That $\hat{f}$ is indeed continuous
follows from 
$\hat{f}(\phi_\alpha)=\phi_\alpha(f)\to\phi(f)=\hat{f}(\phi)$ by 
definition of convergence in $\overline{{\cal U}}$. The isomorphism is 
in fact an isometry since on the one hand
$$
||\hat{f}||:=
\sup_{\phi\in\overline{{\cal U}}}|\hat{f}(\phi)|=
\sup_{\phi\in\overline{{\cal U}}}|\phi(f)|\ge
\sup_{\phi\in{\cal U}}|f(\phi)|=||f||
$$
and on the other hand $|\phi(f)|\le ||f|| \sup_{g\in V,||g||=1}|\phi(g)|
=||f||\; ||\phi||$. But $\phi$ is a character, so $||\phi||=1$ 
(to see this, consider any $f$ with $||f||=1$. Then 
$\phi(|f|^2)=|\phi(f)|^2$ so taking the supremum over all $f$ gives
$||\phi||=||\phi||^2$). Therefore $||\hat{f}||\le ||f||$.
Finally, $\cal U$ is densely embedded into $\overline{{\cal U}}$ by standard
Gel'fand theory.\\
The spectrum is a rather abstract construction. However, just as in the 
case of $\agb$ we can describe it in more intuitive terms.
\begin{Theorem}[Characterization of the spectrum]\label{thc1}
$\overline{\cal U}$ is in one to one correspondence with the set
$\mbox{Fun}(\Sigma,G)$ of $G$-valued functions on $\Sigma$,
the correspondence being given by 
$$
\phi\leftrightarrow U_\phi;\;
(U_\phi(v))_{\mu\nu}=\phi(H_{v,\pi_{fun},\mu,\nu})
$$
where $\pi_{fun}$ denotes the fundamental representation of $G$.
\end{Theorem}
Proof :\\
The proof is pretty trivial :\\
Given $\phi\in\overline{{\cal U}}$ we need to check that 
$\phi(H_{v,\pi_{fun},\mu,\nu})$ are the matrix elements of an element of 
$G$. But this follows from the fact that $\phi$ is a homorphism from the 
algebra of vertex functions into the complex numbers. Therefore, all the 
relations satisfied between elements of that algebra continue to hold 
after mapping with $\phi$. The assertion now follows from the fact that 
$H_{v,\pi_{fun},\mu,\nu}$ satisfies all the properties of the matrix 
elements of an element of $G$.\\
Conversely, given an element $U\in\mbox{Fun}(\Sigma,G)$ we define
$\phi_U\in\overline{{\cal U}}$ by
$\phi_U(H_{v,\pi_{fun},\mu,\nu}):=U_{\mu\nu}(v),\;\phi_U(1):=1$ and 
extend by linearity 
and multiplicativity. Then all we have to show in order to establish that
$\phi_U$ qualifies as an element of the spectrum is that  
$\overline{\phi(f)}=\phi(\overline{f})$. This follows from 
the fact that $G$ is unimodular and unitary, therefore 
$\overline{U(v)}_{\mu\nu}$ is actually a polynomial in $U(v)_{\mu\nu}$.\\
$\Box$\\
Since the spectrum is a compact Hausdorff space, positive linear functionals
$\Gamma$ on $C(\overline{{\cal U}})$ are in one to one correspondence with
regular Borel probability measures on $\overline{{\cal U}}$, the 
correspondence being given by $\Gamma(f)=\int_{\overline{{\cal U}}}d\mu f$.
We will define the measure $\mu_U$ through its corresponding characteristic 
functional $\Gamma_U$.
\begin{Definition} \label{defc2}
\be \label{c2}
\Gamma_U(H_{\vec{v},\vec{\pi},\vec{\mu},\vec{\nu}})=
\left\{ \begin{array}{c}
1 \mbox{ : }H_{\vec{v},\vec{\pi},\vec{\mu},\vec{\nu}}=1\\
0 \mbox{ : otherwise}
\end{array}
\right.
\ee
\end{Definition}
That this completely defines $\Gamma_U$ follows from the fact that finite
linear combinations of vertex functions are dense in $V$ and thus by the 
Gel'fand isomorphism dense in $C(\overline{{\cal U}})$.
It is immediate from this definition that the measure $\mu_U$ is 
diffeomorphism invariant and gauge invariant.
It is easy to check that Definition (\ref{defc2}) is equivalent to the 
following cylindrical definition of $\mu_U$ :\\
A function $f$ of $\phi$ is said to be cylindrical with respect 
to a vertex set $\vec{v}$ if it depends only 
on the finite number of variables $U_\phi(v_1),..,U_\phi(v_n)$, that is,
$f(\phi)=f_{\vec{v}}(U_\phi(v_1),..,U_\phi(v_n))=:
f_{\vec{v}}(p_{\vec{v}}(\phi))$ where $f_{\vec{v}}$ is 
a complex-valued function on $G^n$.
We define the integral of $f$ by 
\be \label{c3}
\int_{\overline{\cal U}} d\mu_U(\phi) f(\phi):=
\int_{\overline{\cal U}} d\mu_{\vec{v}}(\phi) f(\phi):=
\int_{G^n} d\mu_H(g_1)..d\mu_H(g_n) f_{\vec{v}}(g_1,..,g_n))
\ee
where $\mu_H$ is the Haar measure on $G$. The so defined family of measures
$\{\mu_{\vec{v}}\}$ is of course consistently defined because the Haar 
measure is a probability measure. That (\ref{c3}) coincides with 
(\ref{c2}) is now immediate because of the Peter\&Weyl theorem which says 
that the functions $\sqrt{d_\pi}\pi_{ij}(g)$ form an orthonormal basis 
for $L_2(G,d\mu_H)$ : Equation (\ref{c2}) is nothing else than the statement
that the vertex function $1$ is orthogonal to every other one with respect
to $\mu_U$. But this is also the case by (\ref{c3}) using the Peter\&Weyl 
theorem.
\begin{Corollary}
The vertex functions provide an orthonormal basis for 
$L_2(\overline{{\cal U}},d\mu_U)$.
\end{Corollary}
This follows again from the Peter\&Weyl theorem.\\
Another definition of the measure $\mu_U$ is as follows :
\be \label{c4}
d\mu_U(\phi):=\prod_{v\in\Sigma} d\mu_H(U_\phi(v)) .
\ee
Notice that this definition is rigorous : As is shown in \cite{Yamasaki},
a measure which on cylindrical subspaces is a product of probability 
measures corresponds uniquely to a $\sigma$-additive measure on the 
uncountable product measurable space.\\
We also have the analogue of the Marolf-Mour\~ao result \cite{3a} : while 
$\cal U$
is topologically dense in $\overline{{\cal U}}$ it is contained in 
a measurable set of $\mu_U$ measure zero. To see this we show a stronger 
result : the set of $\phi$'s which 
are continuous at any point and in any direction is of measure zero. We will
give a new proof of this result below.
\begin{Theorem} \label{thc2}
Let $p\in\Sigma$ be a point and $c$ a curve in $\Sigma$ starting in $p$.
Let $F_{p,c}:=\{\phi\in\overline{{\cal U}};\; \lim_{t\to 0}
\sum_{\mu\nu}|\phi(H_{c(t),\pi_{fun},\mu,\nu}-H_{p,\pi_{fun},\mu,\nu})|^2
=\lim_{t\to 0}2d(1-\frac{1}{d}\Re\mbox{tr}(U_\phi(c(t))U_\phi(p)^{-1})=0\}$
be the set of characters which are continuous at $p$ along the curve $c$
($d$ is the dimension of the fundamental representation $\pi_{fun}$ of 
$G$).\\ 
Then $\mu_U(F_{p,c})=0$.
\end{Theorem}
Proof :\\
Notice that $F_{p,c}$ is measurable since it can be characterized in 
terms of measurable functions. Consider the measurable function
$f_{\epsilon,t}(\phi):=\exp(-\frac{1}{\epsilon^2}[1-\frac{1}{N}\Re
\mbox{tr}(U_\phi(c(t))U_\phi(p)^{-1})])$. We have $\lim_{\epsilon\to 0}
\lim_{t\to 0}f_{\epsilon,t}(\phi)=\chi_{F_{p,c}}(\phi)$ where $\chi$ denotes
the characteristic function. The two-parameter set of functions 
$f_{\epsilon,t}$ are bounded by the measurable function $1$ and the limit
$\lim_{\epsilon\to 0,t\to 0} f_{\epsilon,t}=\chi_{F_{p,c}}$ exists pointwise.
Thus, since $\mu_U$ is a Borel measure, we may apply the Lebesgue 
dominated convergence theorem to exchange limits and integration so that
$$
\mu_U(F_{p,c})=\lim_{\epsilon\to 0,t\to 0}\int_{\overline{{\cal U}}}
d\mu_U(\phi)f_{\epsilon,t}(\phi)=\lim_{\epsilon\to 0}\int_{G^2} d\mu_H(g) 
d\mu_H(h) e^{-\frac{1}{\epsilon^2}[1-\frac{1}{d}\Re\mbox{tr}(gh^{-1})]}
$$
by the definition of the measure $\mu_U$. We can now see where the measure
zero property has its origin : the measure is diffeomorphism invariant 
and therefore the $t$-dependence has completely dropped out of the integral,
two points are ``arbitrarily far away" as soon as they are different.
To complete the proof use the translation invariance and normalization of 
the Haar measure to see that the last integral becomes
$$\int_G d\mu_H(g) e^{-\frac{1}{\epsilon^2}[1-\frac{1}{N}\Re\mbox{tr}(g)]}
$$
Now as $\epsilon\to 0$ we may expand $g=\exp(t^I\tau_I)$ around $1$ 
which is where 
the main contribution comes from. Then $1-\frac{1}{d}\Re\mbox{tr}(g)=
1/2t^I t^J\delta_{IJ}+o((t^I)^3)$ and we conclude with a similar 
expansion for the Jacobian that the resulting (essntially Gaussian) 
integral is of order $\epsilon^{\dim(G)}$.\\
$\Box$
\begin{Corollary}
Since ${\cal U}\subset F_{p,C}$ for any $p,c$ we have $\mu_U({\cal U})=0$.
\end{Corollary}
Another result that we can copy is the following : the Gel'fand topology
can be characterized by the fact that the (Gel'fand transforms of the)
vertex functions are continuous. Now we can see that the spectrum is 
homeomorphic with the uncountable direct product $X:=\prod_{v\in\Sigma} X_v$
of copies of $G$ (that is, $X_v=G$) equipped with the Tychonov topology.
To see this, recall that the Tychonov topology is the weakest topology 
such that all the projections $p_v\;:\;X\to X_v$ are continuous. Therefore,
every continuous function on $X$ is of the form of sums of 
products of functions of the form $p_v^\star f_v$ where $f_v$ is a 
continuous function on $X_v=G$. But these are precisely the vertex 
functions. Therefore the topologies on $\overline{{\cal U}}$ and $X$ are 
generated by the same functions. That both sets are bijective and inverse 
bijective with each other follows from the fact that the infinite array
of group elements $(g_v)_{v\in\Sigma}$ which is an element of $X$ defines
obviously a $G$-valued function $U(v):=g_v$ and now the assertion 
follows from Theorem \ref{thc1}.\\
This is the precise analogue of the result \cite{3a} that $\agb$ is the 
projective
limit (rather than direct product) of spaces $(\agb)_\gamma$, one for 
each graph $\gamma$, which is isomorphic with $G^n,\;n$ the number of 
edges of $\gamma$.\\
\\
We now wish to show that $\mu_U$ is the unique probability measure on 
$\overline{{\cal U}}$ which incorporates the reality condition that
$\phi_I,p_I$ are real-valued and that realizes the canonical 
commutation relations $[\hat{\phi}_I(x),\hat{p}^J(y)]=i\hbar\delta_I^J
\delta(x,y)$. \\
Notice that for $\phi\in {\cal U}$ whe have that $U_\phi(v)$ is a unitary 
matrix. The operator 
$(\hat{U}_{\mu\nu}(v)f)(\phi):=(U_{\phi}(v))_{\mu\nu} f(\phi)$ is densely
defined on $L_2(\overline{{\cal U}},d\mu_U)$, thus the appropriate 
adjointness relation to be implemented is 
$[\hat{U}_\phi(v)_{\mu\nu}]^\dagger=
[(\hat{U}_\phi(v))^{-1}]_{\nu\mu}$. The operator $\hat{p}^I(x):=-i\hbar
\delta/\delta\phi_I(x)$ on the other hand, which formally implements the 
canonical commutation relations $[\hat{\phi}_I(x),\hat{p}^J(y)]=
i\hbar\delta_I^J\delta(x,y)$, is certainly not densely defined on the $L_2$ 
space. This is 
to be expected because our states depend on $\phi$ {\em at a given 
point}, that is, $\phi$ is not at all smeared with test functions. The 
situation is somewhat similar to the case of $\agb$ where the connection
is only smeared in one direction along the path of the associated 
holonomy : the electric field operators are not densely defined. What 
{\em is} densely defined though is the integral of the electric field over
a two-dimensional surface. The surface and the path form together three
smearing directions which are sufficient to absorb the delta-distribution 
that appears in the canonical commutation relations. Likewise we are lead
here to consider the operators $\hat{p}^I(B):=\int_B d^3x \hat{p}^I(x)$ 
which gives the required three-dimensional smearing over a three-dimensional 
compact subset $B\subset\Sigma$. We ask that $[\hat{p}^I(B)]^\dagger=
\hat{p}^I(B)$ for each $B$. Notice that the integral is covariant under 
diffeomorphisms because $p_I$ is a density of weight one.\\
Let us check that $\hat{p}^I(B)$ is indeed densely defined. Let
$f_{\vec{v}}$ be a cylindrical function. We are going to evaluate
$\hat{p}^I(B)$ on $f_{\vec{v}}$ first of all for a smooth field $\phi$
and then are able to generalize to $\overline{{\cal U}}$. We have
formally
\be \label{c5}
\hat{p}^I(B)f_{\vec{v}}(\phi)=-i\hbar\int_B d^3x  
\sum_{v\in\vec{v}}\frac{\delta U_{v,\mu\nu}(\phi)}{\delta\phi_I(x)}
\frac{\partial f_{\vec{v}}(\phi)}{\partial U_{v,\mu\nu}(\phi)}
\ee
and everything boils down to the appropriate definition of the functional
derivative involved in (\ref{c5}). As it stands, it is ill-defined.
We will proceed as follows : let $h_\epsilon(x,y)$ be a one-parameter 
family of smooth functions such that 
$\lim_{\epsilon\to 0}\int_\Sigma d^3y f(y) h_\epsilon(x,y)=f(x)$ for every
smooth function, say, of rapid decrease. Let $\alpha_{r,s}$ be a two 
parameter family of smooth loops which embed an abstract solid cylinder into
$\Sigma$ with $r,s\in [0,1]$ and such that $v$ lies inside the image
of $[0,1]^2\times S^1$. It follows that the equation $\alpha_{r,s}(t)=v$
has only one solution in $[0,1]^3$.\\
Consider the regulated quantity
\ba \label{c6}
U^\epsilon_v(\phi)&:=&{\cal P}\exp(\int_0^1 dt \phi(\alpha,\epsilon,v,t))
\mbox{ with } 
\nonumber\\
\phi(\alpha,\epsilon,v,t) &:=&\int_0^1 dr\int_0^1 ds
h_\epsilon(\alpha_{r,s}(t),v)
|\det(\frac{\partial\alpha_{rs}(t)}{\partial(r,s,t)})|
\phi_I(\alpha_{r,s}(t))\tau_I\;. 
\ea
where $\cal P$ means path ordering. Notice that $\lim_{\epsilon\to 0}
U^\epsilon_v(\phi)=U_v(\phi)$. There are two reasons for why we chose the 
particular smearing (\ref{c6}) of $U_v(\phi)$ which involves the path 
ordering :\\ 
The first motivation is 
that the point holonomies $U_v(\phi)$ are in a certain sense limits of 
usual loop holonomies as the the loop shrinks to a point, evaluated on a 
distributional connection (for a smooth connection the limit would equal
$1_G$). Indeed, for each loop $\alpha_{r,s}(t)$ we may associate a
G-``connection" 
$$\un{A}_a(\alpha_{rs}(t)):=h_\epsilon(\alpha_{r,s}(t),v)
|\det(\frac{\partial\alpha_{rs}(t)}{\partial(r,s,t)})|
\phi_I(\alpha_{r,s}(t))\tau_I
\frac{\dot{\alpha}^a_{rs}(t)}{||\dot{\alpha}_{rs}(t)||^2} 
$$
where we have used a Euclidean norm $||.||$ to raise indices. The meaning
of this quantity is that $\int dr\int ds\oint_{\alpha_{rs}}\un{A}=
\int dt \phi(\alpha,\epsilon,v,t)$.\\
The second motivation is that the naive definition
$\delta U_v(\phi)/\delta\phi_I(x):=\delta(x,v)\partial U_v(\phi)/\partial
\phi_I(v)$, when integrated over $B$, does not result in a function which
can be expressed as a polynomial in $U_{v,\mu\nu}$ again so that 
$p^I(B)$ would not be densely defined. The particular choice (\ref{c6})
will lead to a densely defined operator.

The functional derivative of the quantity (\ref{c6}) is well-defined :
We get just zero unless there is a solution 
$\alpha_{r_x,s_x}(t_x)=x$ with $(r_x,s_x,t_x)\in (0,1)^2\times S^1$ in which 
case we find 
\ba \label{c7}
&&\frac{\delta U^\epsilon_v(\phi)}{\delta \phi_I(x)}
\nonumber\\
&=& \int_{[0,1]^3}dr ds dt \delta(x,\alpha_{rs}(t))
h_\epsilon(\alpha_{r,s}(t),v)
|\det(\frac{\partial\alpha_{rs}(t)}{\partial(r,s,t)})|
U^\epsilon_v(0,t_x)\tau_I U^\epsilon_v(t_x,1)
\nonumber\\
&=& h_\epsilon(x,v)
\int_{[0,1]^3}dr ds dt \delta(r,r_x)\delta(s,s_x)
\frac{1}{2}[\delta(t,t_{x+})+\delta(t,t_{x-})]\times\nonumber\\
&\times&
\frac{|\det(\frac{\partial\alpha_{rs}(t)}{\partial(r,s,t)})|}
{|\det(\frac{\partial\alpha_{rs}(t)}{\partial(r,s,t)})_{r_x,s_x,t_x}|}
U^\epsilon_v(0,t_x)\tau_I U^\epsilon_v(t_x,1)
\nonumber\\
&=& h_\epsilon(x,v)
\frac{1}{2}[U^\epsilon_v(0,t_x+)\tau_I U^\epsilon_v(t_x+,1)
+U^\epsilon_v(0,t_x-)\tau_I U^\epsilon_v(t_x-,1)]\;.
\ea
Here we mean by $U^\epsilon_v(a,b)$ the path ordered exponential 
(\ref{c6}) for $t\in [a,b]$ where we identify values modulo $1$.
Now we easily see that $\lim_{\epsilon\to 0}U^\epsilon_v(t_x,t_x+1)$
gives $U(v)(\phi)$ if $t_x=0$ and $1_G$ otherwise. However, since 
$h_\epsilon$ gives zero as $\epsilon\to 0$ unless $x=v$ for which
indeed $t_x=0\equiv 1(\mbox{mod } 1)$ we find the 
functional derivative of $U_v(\phi)$ to be 
\be \label{c9}
\frac{\delta U_{v\mu\nu}(\phi)}{\delta\phi_I(x)}:=
\lim_{\epsilon\to 0}\frac{\delta U^\epsilon_{v\mu\nu}(\phi)}{\delta\phi_I(x)}
=\delta(x,v)
\frac{1}{2}[U_v(\phi)\tau_I+\tau_I U_v(\phi)]_{\mu\nu} \;.
\ee
The end result (\ref{c9}) is independent of the particular function
$h_\epsilon$ or family of loops $\alpha_{r,s}$ of loops used and is therefore
satisfactory.\\
Putting everything together we finally find the remarkably simple formula
\be \label{c10}
p^I(B)f_{\vec{v}}=-i\hbar\sum_{v\in B\cap\vec{v}} X^I_v f_{\vec{v}}
\ee
where
\ba \label{c11}
X^I_v &:=& X^I(U_v(\phi)),\;X^I(g):=\frac{1}{2}[X^I_R(g)+X^I_L(g)],
\nonumber\\
X^I_R(g)&:=&\mbox{tr}([\tau_I g]^T\frac{\partial}{\partial g}),
\nonumber\\
X^I_L(g)&:=&\mbox{tr}([g\tau_I]^T\frac{\partial}{\partial g}).
\ea
The vector fields $X^I_R$ and $X^I_L$ on $G$ are respectively right and left
invariant in the sense $X_R(gh)=X_R(g),\;X_L(hg)=X_L(g)$ for each
$h\in G$ as one can easily check. However, 
$X^I_R(hg)=\mbox{Ad}(h^{-1})^{IJ} X^J_R(g),\;X_L(gh)=\mbox{Ad}(h)^{IJ}
X^J_L(g)$ where $h\tau^I h^{-1}=\mbox{Ad}(h)^{IJ} \tau^J$ denotes the 
matrix elements of the adjoint representation of $G$ on $Lie(G)$. This
is an important consistency check because it implies that 
$X^I(hgh^{-1})=Ad(h)^{IJ}X^J(g)$ transforms according to adjoint 
representation as well and so (\ref{c10}) is covariant under gauge 
transformations. \\
Another intuively appealing feature of (\ref{c10}) is the following :
recall that the functional derivative of a usual path holonomy $h_e$ at $x$
is proportional to a right invariant vector field $X_R(h_e)$ if 
$x$ is 
the starting point of $e$ and proportional, with the {\em same} 
coefficient, to a left invariant vector field $X_L(h_e)$ 
if $x$ is the end point of $e$. Therefore, formally, we expect that
if $e$ shrinks to a point (or if $e$ is a loop) then we must get the 
average $X(h_e)=(X_R(h_e)+X_L(h_e))/2$ which is exactly the result 
(\ref{c10}), thus
reassuring that our interpretation of $U(v)$ as a point holonomy is 
correct.\\
As one can easily check, if $[\tau_I,\tau_J]=f_{IJK} \tau_K$
define the structure constants of $Lie(G)$ 
then $[X^I_R,X^J_R]=-f_{IJK}X^K_R,\; [X^I_L,X^J_L]=+f_{IJK}X^K_L,
[X^I_R,X^J_L]=0$ which implies that 
$[X^I,X^J]=f_{IJK}(-X^K_R+X^K_L)/4$ so that the algebra of the $X^I$
is not closed while the algebra of the $(-X_R+X_L)/2$ is closed. We note 
that if $X^I_\pm=\frac{1}{2}[X^I_L\pm X^I_R]$ then 
$[X^I_+,X^J_+]=[X^I_-,X^J_-]=\frac{f_{IJK}}{4}X^K_-,\;
[X^I_+,X^J_-]=\frac{f_{IJK}}{4}X^K_+$. That we do not find the right hand 
side of $[p^I(B),p^J(B)]$ to be expressible in terms of a $p^I(B')$
again is not surprising as the Poisson bracket between the corresponding
functions should actually vanish. The reason for this ``anomaly" is due to 
the fact that we did not properly smear the $\phi$'s as explained in 
\cite{ACLZ} : we can expect the correspondence $\{.,.\}\to\frac{1}{i\hbar}
[.,.]$ to hold {\em everywhere} only if $p$'s and $\phi$'s are smeared 
in three directions. Otherwise even classically we get in conflict because
if one assumes $\{p^I(B),p^J(B')\}=0$ then the Jacobi ``identity" is 
violated. It is violated precisely because the identity holds only on 
twice (functionally) differentiable functions which is not the case here.
We conclude that nothing is inconsistent. See \cite{ACLZ} for a more
detailed explanation for this phenomenon.

Interestingly, we do not encounter any such ``anomaly" if we restrict 
attention to gauge invariant functions of $p$ which are the only 
ones in which we are interested in anyway : indeed, as we saw the ``anomaly"
arose only due to the fact that the $X^I$ do not commute which suggests 
to build gauge invariant quantities. In fact, classically only the modulus
of the vector valued density $p^I$ is gauge invariant so that we are lead
to consider the quantity $\hat{K}(B):=\hat{p}^I(B)\hat{p}^J(B)\delta_{IJ}$
which on vertex functions gives $\hat{K}(B)=-\hbar^2\sum_{v\in B\cap\vec{v}}
\delta_{IJ}X^I_v X^J_{v'}f_{\vec{v}}$. Choosing $B=B_v$ small enough a 
neighbourhood of a point $v$ we find that 
$\hat{K}(B_v)=:\hat{K}_v^2=-\hbar^2 
X^I_v X^J_v \delta_{IJ}$ which is gauge invariant and all the 
$\hat{K}_v$ actually commute as they should.

In any case we are now in the position to show that the adjointness
relations $\hat{U}_{v,\mu\nu}(\phi)^\dagger=
\hat{U}^{-1}_{v,\nu\mu}(\phi),\;(p^I(B))^\dagger=p^I(B)\forall v,B$ hold
on ${\cal H}=L_2(\overline{{\cal U}},d\mu_U)$. The first relation follows
trivially from the fact that $\hat{U}_{v,\mu\nu}(\phi)$ is a
multiplication operator on $\cal H$ and so its hermitean conjugate is its
complex conjugate.\\ The second follows from the fact that $X^I_v$ is a
linear combination of left and right invariant vector fields on $G$ both
of which annihilate the volume form corresponding to the (left and right
invariant) Haar measure. Moreover, by results following from \cite{2} one
can show that $\hat{p}^I(B)$ is essentially self-adjoint and 
that the adjointness relation on $\hat{U}_{v,\mu,\nu}$ is identically 
satisfied.\\ 
This suffices to show that $\mu_U$
implements the $^\star$ relations. Now, since $p^I$ is linear in left and
right invariant vector fields and $\hat{U}_{v,\mu\nu}$ is a multiplication
operator, the measure $\mu_U$ is {\em uniquely} selected (if we insist
that it be a probability measure) by the above reality conditions as was
shown for a general theory in \cite{4} simply because the Haar measure on
a compact gauge group is by definition annihilated by precisely the left
and right invariant vector fields.\\ 
\\ 
We now address the diffeomorphism invariance.\\ 
The space $\Phi_U$ is the space of finite linear
combinations of vertex functions equipped with the ``Fourier topology"
(similar as in the case of the fermion field)
$$
||f||_1:=\sum_{\vec{v},\vec{\pi},\vec{\mu},\vec{\nu}}
|<H_{\vec{v},\vec{\pi},\vec{\mu},\vec{\nu}},f>| $$ and $\Phi'_U$ its
topological dual (alternatively, as in \cite{4}, we could equip $\Phi_U$
with a standard nuclear topology on $G^n$ where $n$ is the number of
vertices of the cylindrical function in question). Again we have the
inclusion $\Phi_U\subset{\cal H}_U\subset\Phi'_U$ and we are looking for
solutions of the diffeomorphism constraint in $\Phi'_U$ by a
straightforward application of the group averaging method from \cite{4}.
The unitary representation of the diffeomorphism group is defined by
$\hat{U}(\varphi)f_{\vec{v}}:=f_{\vec{\varphi(v)}}$ for each smooth
diffeomorphism $\varphi$ of $\Sigma$ (in case that $\vec{v}$ are vertices
of a graph $\gamma$ defined by gauge fields as below consider only those
smooth diffeomomorphisms that leave $\gamma$ piecewise analytic). Let us
use a compound label $I$ for $(\vec{\pi},\vec{\mu},\vec{\nu})$. Consider
the orbit $\{H_{\vec{v},I}\}:=\{H_{\vec{\varphi(v)},I},\; \varphi\in
\mbox{Diff}(\Sigma)\}$. Then we define 
$$
[H_{\vec{v},I}]:=\sum_{H\in\{H_{\vec{v},I}\}}H 
$$ and extend by linearity,
that is, decompose every $f\in\Phi_U$ first in terms of $H_{\vec{v},I}$'s
and then average each $T_{\vec{v},I}$ separately.
We quickly verify that this method always results in elements of $\Phi'_U$. 
The diffeomorphism invariant Hilbert space is then defined through
$<[f],[g]>_{Diff}:=[f](g)$ for any $f,g\in \Phi_U$.

\section{Gauge Invariant States}

Let us summarize : our Hilbert space of diffeomorphism non-invariant
(not necessarily gauge invariant)
functions of gravitational, gauge, spinor and Higgs fields is given by
$${\cal H}:=
L_2(\Ab_{SU(2)},d\mu_{AL}(SU(2)))\otimes
L_2(\Ab_G,d\mu_{AL}(G))\otimes L_2(\overline{{\cal S}},d\mu_F)
\otimes L_2(\overline{{\cal U}},d\mu_U). $$

The Hilbert space of gauge invariant functions will be just the restriction
of $\cal H$ to gauge invariant functions. It is easy to see that, 
because 
our total measure is a probability measure, gauge invariant functions 
will be still integrable with respect to it.

A natural gauge invariant object associated with spinor fields, Higgs 
fields and gauge 
fields are ``spin-colour-network states". By this we mean the following :
Let $\gamma$ be a piecewise analytic graph with edges $e$ and vertices $v$
which is not necessarily connected or closed.
By suitably subdividing edges into two halves we can assume that all edges
are outgoing at a vertex.
Given a (generalized) connection $\omega_a=A_a+\underline{A}_a$
we can compute the holonomies 
$h_e(A),\;\underline{h}_e(\underline{A}),H_e(\omega)=h_e(A)\underline{h}_e(
\underline{A})$. With each edge $e$ we associate a spin $j_e$ and a
colour $c_e$ corresponding to irreducible representations of $SU(2)$ and 
of $G$ respectively (for instance for $G=SU(N)$, $c_e$ is an array
of $N-1$ not increasing integers corresponding to the frame of a 
Young diagramme). Furthermore, with each vertex $v$ we associate an integer
$0\le n_v\le 2d$, yet another colour $C_v$ and two projectors $p_v,q_v$. The 
integer $n_v$ corresponds to the subvector space of $Q_v$,
the vector space spanned by holomorphic functions of 
$\theta_i(v)$, spanned by those vectors $F_{I,v}$ such that $|I|=n(v)$.
Likewise, the colour $C_v$
stands for an irreducible representation of $G$, evaluated at the point 
holonomy $U(v)$. 
The projector $p_v$ is a certain $SU(2)$ invariant matrix which projects 
onto one of the linearly independent 
trivial representations contained in the decomposition into irreducibles
of the tensor product consisting of\\
a) the $n_v-$fold tensor product 
of fundamental representations of $SU(2)$ associated with the 
subvector space of $Q_v$ spanned by the $F_{I,v},\;|I|=n(v)$ and\\
b) the tensor product of the
irreducible representations $j_e$ of $SU(2)$ of spin $j_e$ where $e$ runs 
through the subset of edges of $\gamma$ which start at $v$.\\
Likewise,
the projector $q_v$, repeats the same procedure just that $SU(2)$
is being replaced by $G$ and that we need to consider in addition the 
adjoint representation associated with $C_v$ coming from the Higgs field at
$v$. 
Now we simply contract all the indices of the tensor product of\\ 
1) the irreducible representations evaluated at the holonomy of the given 
connection, \\
2) the fundamental representations evaluated at the given spinor field 
and\\ 
3) the adjoint representations evaluated at the given scalar field,\\
all associated with the same vertex $v$, with the projectors $p_v,q_v$ in 
the obvious way and for all $v\in V(\gamma)$. The result is a 
gauge invariant state 
$$ T_{\gamma,[\vec{j},\vec{n},\vec{p}],[\vec{c},\vec{C},\vec{q}]}$$
which we will call a spin-colour-network state because they extend the 
definition of the pure spin-network states which arise in the source-free
case (e.g. \cite{4}).\\
These spin-colour-networks turn out to be a basis for the subspace of 
gauge invariant functions. They are not orthonormal, but almost : we just
need to decompose the fermionic dependence into an orthonormal basis 
of the $Q_v$.\\ 
This means that we get a gauge invariant Hilbert space
$${\cal H}_{inv}:=
L_2([\Ab_{SU(2)}\times\Ab_G\times\overline{{\cal S}}
\times\overline{{\cal U}}]/\overline{{\cal G}},
d\mu_{AL}(SU(2))\otimes d\mu_{AL}(G)\otimes d\mu_F\otimes d\mu_U) 
$$
where $\overline{{\cal G}}$ denotes the action of the gauge group 
$SU(2)\times G$ on all the fields at each point of $\Sigma$.\\ 
\\
Next, in order to get the full space of diffeomorphism invariant
distributions consider in addition to the spaces 
$\Phi_F,\Phi_U$ the spaces $\Phi_{SU(2)},\Phi_G$ of gravitational 
and gauge field finite linear combinations of not necessarily gauge invariant
(generalized) spin-network and colour-network functions \cite{4} equipped 
with a nuclear topology from $SU(2)^n$ and $G^n$ respectively, $n$ the 
number of edges of the graph of the cylindrical function in question.
Let $\Phi'_{SU(2)},\Phi'_G$ be their topological duals. We now consider 
the product topological vector space and its dual
$$
\Phi=\Phi_{SU(2)}\times\Phi_G\times\Phi_F\times\Phi_U \mbox{ and }
\Phi'=\Phi'_{SU(2)}\times\Phi'_G\times\Phi'_F\times\Phi'_U 
$$ 
take the gauge invariant subspaces thereof 
and construct diffeomorphism invariant elements of $\Phi'$ as follows :
take a spin-colour-network function $T$, consider its orbit under analyticity
preserving diffeomorphisms $\{T\}$ and define $[T]:=\sum_{T'\in\{T\}} T'$
(slightly modified as in \cite{6a} through the presence of the graph 
symmetries). Take any $f\in\Phi$, decompose it in terms of $T$'s,
average each $T$ separately and define the result to be $[f]$.
Then $<[f],[g]>_{Diff}:=[f](g)$ for each $f,g\in\Phi$ defines the Hilbert 
space ${\cal H}_{Diff}$ upon completion.
\\
\\
\\ 
\\ 
{\large Acknowledgements}\\
\\
This research project was supported in part by DOE-Grant
DE-FG02-94ER25228 to Harvard University.


\begin{thebibliography}{99}

\parskip -5pt


\bibitem{Jaffe} J. Glimm and A. Jaffe, ``Quantum physics", 
Springer-Verlag, New York, 1987

\bibitem{Yamasaki} Y. Yamasaki, ``Measures on Infinite Dimensional Spaces",
World Scientific, Singapore, 1984

\bibitem{4} A. Ashtekar, J. Lewandowski, D. Marolf, J. Mour\~ao, T.
Thiemann, ``Quantization for diffeomorphism invariant theories 
of connections with local degrees of freedom", Journ. Math. Phys.
{\bf 36} (1995) 519-551

\bibitem{0} A. Ashtekar and C.J. Isham,
Class. \& Quan. Grav. {\bf 9}, 1433 (1992).

\bibitem{1} A. Ashtekar and J. Lewandowski, ``Representation
theory of analytic holonomy $C^\star$ algebras'', in {\it Knots and
quantum gravity}, J. Baez (ed), (Oxford University Press, Oxford 1994).

\bibitem{2} A. Ashtekar and J. Lewandowski, ``Differential
geometry on the space of connections via graphs and projective
limits'', Journ. Geo. Physics {\bf 17} (1995) 191

\bibitem{3} A. Ashtekar and J. Lewandowski, J. Math. Phys. {\bf 36}, 2170
(1995). 

\bibitem{3a} D. Marolf and J. M. Mour\~ao, ``On the support of the
Ashtekar-Lewandowski measure'',  Commun. Math. Phys. {\bf 170} (1995)
583-606

\bibitem{allmt1} A. Ashtekar, J. Lewandowski, D. Marolf, J. Mour\~ao
and T. Thiemann, ``A manifestly gauge invariant approach to quantum
theories of gauge fields'', in {\it Geometry of constrained dynamical
systems}, J. Charap (ed) (Cambridge University Press, Cambridge,
1994); ``Constructive quantum gauge field theory in two space-time
dimensions'' (CGPG preprint).

\bibitem{almmt2} A. Ashtekar, J. Lewandowski, D. Marolf, J. Mour\~ao,
T. Thiemann,
``$SU(N)$ Yang-Mills theory in two dimensions : A complete solution",
Preprints CGPG-95/4-2, HUTMP-96/B-351, hep-th/9605128


\bibitem{Higuchi} A. Higuchi Class. Quant. Grav. {\bf 8} (1991) 1983 \\
                  A. Higuchi Class. Quant. Grav. {\bf 8} (1991) 2023


\bibitem{Marolf} D. Marolf, ``The spectral analysis inner product for
quantum gravity,'' preprint gr-qc/9409036, to appear in the
Proceedings of the VIIth Marcel-Grossman Conference, R. Ruffini and
M. Keiser (eds) (World Scientific, Singapore, 1995); D. Marolf,
Ph.D. Dissertation, The University of Texas at Austin (1992).\\
D. Marolf, ``Quantum observable and recollapsing
dynamics,'' preprint gr-qc/9404053. Class. Quant. Grav. {\bf 12} (1995) 
1199\\
D. Marolf ``Almost Ideal Clocks in Quantum Cosmology: A Brief
Derivation of Time,'' preprint gr-qc/9412016.

\bibitem{AA} A.\ Ashtekar, Phys.\ Rev. Lett.\ {\bf 57} 2244 (1986),
            Phys.\ Rev.\ {\bf D36}, 1587 (1987).

\bibitem{Barbero} F. Barbero, Phys. Rev. D {\bf 51} (1995) 5507

\bibitem{RM-T} H. Morales-Tec\'otl, C. Rovelli, Phys. Rev. Lett. {\bf 72}
(1995) 3642-3645 \\
H. Morales-Tec\'otl, C. Rovelli, Nucl. Phys. {\bf B331} (1995) 325-361

\bibitem{5} T. Thiemann, ``Anomaly-free formulation of non-perturbative,
four-dimensional, Lorentzian quantum gravity", Harvard University
Preprint HUTMP-96/B-350, gr-qc/960688, Physics Lett. N {\bf 380} (1996)
257-264

\bibitem{6} T. Thiemann, ``Quantum Spin Dynamics (QSD)", Harvard University 
Preprint HUTMP-96/B-359, gr-qc/96066089\\
            T. Thiemann, ``Quantum Spin Dynamics (QSD) II : The Kernel of 
the Wheeler-DeWitt Constraint Operator" , Harvard University Preprint 
HUTMP-96/B-352, gr-qc/9606090

\bibitem{6a} T. Thiemann, ``QSD III : Quantum Constraint Algebra and 
Physical Scalar Product in Quantum General Relativity", Harvard University
Preprint HUTMP-97/B-363

\bibitem{6b} T. Thiemann, ``QSD IV : 2+1 Euclidean Quantum Gravity as a 
model to test 3+1 Lorentzian Quantum Gravity", Harvard University Preprint
HUTMP-96/B-360

\bibitem{6c} T. Thiemann, ``QSD VI : Quantum Poincar\'e Algebra and a 
Quantum Positivity of Energy Theorem for Canonical Quantum Gravity", Harvard
University Preprint HUTMP-97/B-356

\bibitem{BK} J. Baez, K. Krasnov, ``Quantization of diffeomorphism invariant
theories with fermions", hep-th/9703112

\bibitem{6d} T. Thiemann, ``QSD V : Quantum Gravity as the Natural Regulator
of Matter Quantum Field Theories", Harvard University Preprint 
HUTMP-96/B-357

\bibitem{Jacobson} T. Jacobson, ``Fermions in canonical gravity",
Class. Quantum Grav. {\bf 5} (1988) L143-148

\bibitem{Berezin} Y. Choquet-Bruhat, C. DeWitt-Morette, ``Analysis, 
Manifolds and Physics Part II : 92 Applications", North Holland, 
Amsterdam, 1989

\bibitem{DeWitt} B. DeWitt, ``Supermanifols", Cambridge University Press,
Cambridge, 1992


\bibitem{ACLZ} A. Ashtekar, A. Corichi, J. Lewandowski, J.-A. Zapata 
``Quantum Theory of Geometry II : Non-commutativity of Riemannian 
Structures", 1997, (in preparation)



\end{thebibliography}
\end{document}